\documentclass[]{llncs}

\usepackage{bussproofs}
\usepackage{listings}
\usepackage{fancybox}
\usepackage{mathtools}
\usepackage{tikz}
\usetikzlibrary{arrows.meta, shapes, arrows, automata, fit, matrix,
  positioning, decorations.pathreplacing, calligraphy}
\usepackage{amssymb}
\usepackage{amsmath}
\usepackage{ifthen}
\usepackage{thm-restate}
\usepackage{caption}
\usepackage{subcaption}
\usepackage{paralist}
\usepackage[hidelinks]{hyperref}

\usepackage[draft]{commenting}

\declareauthor{ri}{Radu}{blue}
\declareauthor{mb}{Marius}{red}
\declareauthor{lb}{Lucas}{green}

\newif\ifLongVersion\LongVersiontrue

\ifLongVersion
\usepackage[createShortEnv,conf={normal}]{proof-at-the-end}
\else
\usepackage[createShortEnv,conf={end,no link to proof,restate}]{proof-at-the-end}
\fi

\ifLongVersion
\newenvironment{myLemmaE}{\begin{lemmaE}}{\end{lemmaE}}
\newenvironment{myTextE}{}{}
\else

\fi

\newcommand{\nat}{\mathbb{N}}
\newcommand{\qpf}{$\mathrm{qpf}$}
\newcommand{\arityof}[1]{\#{#1}}
\newcommand{\lenof}[1]{|{#1}|}
\newcommand{\sizeof}[1]{\mathrm{size}({#1})}
\newcommand{\maxarityof}[1]{{\#}({#1})}
\newcommand{\maxinterof}[1]{\mathsf{N}({#1})}
\newcommand{\maxpredsof}[1]{\mathsf{H}({#1})}

\newcommand{\universe}{\mathbb{C}}
\newcommand{\vars}{\mathbb{V}}

\newcommand{\preds}{\mathbb{A}}
\newcommand{\isdef}{\stackrel{\scriptscriptstyle{\mathsf{def}}}{=}}
\newcommand{\interv}[2]{[{#1},{#2}]}
\newcommand{\tuple}[1]{\langle {#1} \rangle}
\newcommand{\Tuple}[1]{\left\langle {#1} \right\rangle}
\newcommand{\set}[1]{\{ {#1} \}}
\newcommand{\Set}[1]{\left\{ {#1} \right\}}
\newcommand{\pow}[1]{\mathrm{pow}({#1})}

\newcommand{\dom}[1]{\mathrm{dom}({#1})}

\newcommand{\cardof}[1]{|\!| {#1} |\!|}

\newcommand{\finsubseteq}{\subseteq_{\mathit{fin}}}

\newcommand{\bigO}{\mathcal{O}}
\newcommand{\twoexptime}{$2\mathsf{EXP}$}

\newcommand{\fourexptime}{$4\mathsf{EXP}$}
\newcommand{\fiveexptime}{$5\mathsf{EXP}$}

\newcommand{\conp}{$\mathsf{co}$-$\mathsf{NP}$}
\newcommand{\exptime}{$\mathsf{EXP}$}

\newcommand{\comps}{\mathcal{C}}

\newcommand{\interacs}{\mathcal{I}}

\newcommand{\interac}[4]{\tuple{{#1}.{\mathit{#2}}, {#3}.{\mathit{#4}}}}

\newcommand{\intertypeof}[1]{\mathsf{I}({#1})}
\newcommand{\intertype}{\tau}

\newcommand{\arity}{\#}

\newcommand{\beh}{\mathbb{B}}

\newcommand{\states}{\mathcal{Q}}

\newcommand{\ports}{\mathcal{P}}

\newcommand{\arrow}[2]{\xrightarrow{{\scriptstyle #1}}_{{\scriptstyle #2}}}

\makeatletter
\newcommand*{\da@rightarrow}{\mathchar"0\hexnumber@\symAMSa 4B }
\newcommand*{\da@leftarrow}{\mathchar"0\hexnumber@\symAMSa 4C }
\newcommand*{\xdashrightarrow}[2][]{%
  \mathrel{%
    \mathpalette{\da@xarrow{#1}{#2}{}\da@rightarrow{\,}{}}{}%
  }%
}
\newcommand{\xdashleftarrow}[2][]{%
  \mathrel{%
    \mathpalette{\da@xarrow{#1}{#2}\da@leftarrow{}{}{\,}}{}%
  }%
}
\newcommand*{\da@xarrow}[7]{%
  \sbox0{$\ifx#7\scriptstyle\scriptscriptstyle\else\scriptstyle\fi#5#1#6\m@th$}%
  \sbox2{$\ifx#7\scriptstyle\scriptscriptstyle\else\scriptstyle\fi#5#2#6\m@th$}%
  \sbox4{$#7\dabar@\m@th$}%
  \dimen@=\wd0 %
  \ifdim\wd2 >\dimen@
    \dimen@=\wd2 %
  \fi
  \count@=2 %
  \def\da@bars{\dabar@\dabar@}%
  \@whiledim\count@\wd4<\dimen@\do{%
    \advance\count@\@ne
    \expandafter\def\expandafter\da@bars\expandafter{%
      \da@bars
      \dabar@ 
    }%
  }%
  \mathrel{#3}%
  \mathrel{%
    \mathop{\da@bars}\limits
    \ifx\\#1\\%
    \else
      _{\copy0}%
    \fi
    \ifx\\#2\\%
    \else
      ^{\copy2}%
    \fi
  }%
  \mathrel{#4}%
}
\makeatother

\newcommand{\store}{\nu}

\newcommand{\statemap}{\varrho}
\newcommand{\config}{(\comps,\interacs,\statemap)}
\newcommand{\aconfig}{\gamma}

\newcommand{\configset}{\Gamma}

\newcommand{\step}[1]{\xRightarrow{#1}{}}
\newcommand{\Step}{\leadsto}
\newcommand{\Havoc}{\Step^{*}}

\newcommand{\degreenode}[2]{\delta_{#1}({#2})}
\newcommand{\degreeof}[1]{\degreenode{}{#1}}

\newcommand{\comp}{\bullet}

\newcommand{\cl}{\textsf{CL}}
\newcommand{\predname}[1]{\mathsf{#1}}
\newcommand{\apred}{\predname{A}}
\newcommand{\bpred}{\predname{B}}

\newcommand{\emp}{\predname{emp}}

\let\Asterisk\undefined
\newcommand{\Asterisk}{\mathop{\scalebox{1.9}{\raisebox{-0.2ex}{$\ast$}}}\hspace*{1pt}}%
\renewcommand{\vec}[1]{\mathbf #1}
\newcommand{\fv}[1]{\mathrm{fv}({#1})}

\newcommand{\compin}[2]{{#1}@{#2}}
\newcommand{\compact}[1]{[{#1}]}
\newcommand{\compactin}[2]{\compin{\compact{#1}}{#2}}
\newcommand{\interacn}[4]{\tuple{{#1}.\mathit{#2}, \ldots, {#3}.\mathit{#4}}}
\newcommand{\interactwo}[4]{\tuple{{#1}.\mathit{#2}, {#3}.\mathit{#4}}}

\newcommand{\finmap}{\rightharpoonup_{\scriptscriptstyle\mathit{fin}}}

\newcommand{\profile}[1]{\lambda_{\scriptscriptstyle #1}}

\newcommand{\asid}{\Delta}

\newcommand{\arule}{\mathsf{r}}
\newcommand{\ring}[2]{\predname{ring}_{{#1},{#2}}}
\newcommand{\pcring}[2]{\overline{\predname{ring}}_{{#1},{#2}}}
\newcommand{\chain}[2]{\predname{chain}_{{#1},{#2}}}

\newcommand{\size}[1]{\mathrm{size}({#1})}

\newcommand{\defn}[2]{\mathrm{def}_{#1}({#2})}

\newcommand{\defnof}[1]{\mathrm{Def}({#1})}

\renewcommand{\mod}{~\mathrm{mod}~}

\newcommand{\hoare}[3]{\{ {#1} \} ~\mathsf{#2}~ \{ {#3} \}}

\newcommand{\hinv}[2]{\mathsf{HavocInv}[{#1},{#2}]}
\newcommand{\sinv}[2]{\mathsf{StepInv}[{#1},{#2}]}
\newcommand{\bnd}[2]{\mathsf{DegreeBound}[{#1},{#2}]}

\newcommand{\lang}[1]{\mathcal{L}({#1})}
\newcommand{\langof}[2]{\mathcal{L}_{#1}({#2})}

\newcommand{\toktoken}{\mathsf{T}}
\newcommand{\toknotok}{\mathsf{H}}

\newcommand{\treenode}{\mathit{Node}}
\newcommand{\treeleaf}{\mathit{Leaf}}
\newcommand{\treeroot}{\mathit{Root}}
\newcommand{\req}{\mathit{req}}
\newcommand{\reply}{\mathit{reply}}
\newcommand{\inp}{\mathit{in}}
\newcommand{\outp}{\mathit{out}}

\newcommand{\codesize}{\footnotesize}

\newcounter{index}

\newcommand{\tree}{\mathfrak{t}}
\newcommand{\treesof}[1]{\mathbb{T}({#1})}
\newcommand{\subtree}[2]{{#1}_{|{#2}}}

\newcommand{\auth}{\mathcal{A}}
\newcommand{\talpha}{\Sigma}
\newcommand{\tastates}{\mathcal{S}}
\newcommand{\tafinstates}{\mathcal{F}}
\newcommand{\tatrans}{\delta}
\newcommand{\trans}{\mathcal{T}}
\newcommand{\run}{\pi}
\newcommand{\proj}[2]{{#2}{\downarrow_{#1}}}
\newcommand{\cyl}[2]{{#2}{\uparrow^{#1}}}

\newcommand{\substitution}[3]{ {[ #1 / #2] }_{#3}}

\newcommand{\ruleform}{\ensuremath{ 
	\apred(x_1, \ldots, x_{\arityof{\apred}}) \leftarrow 
	\exists y_1 \ldots \exists y_m ~.~ \phi * \Asterisk_{\ell\in\interv{1}{h}} 
	~\bpred_\ell(z^\ell_1, \ldots, z^\ell_{\arityof{\bpred_\ell}})}}

\newcommand{\tastatepred}[1]{\ensuremath{ q_{#1} }} 

\newcommand{\tavarin}[1]{\ensuremath{ {\widetilde{x}}_{#1} }}
\newcommand{\tavarout}[2]{\ensuremath{ {\widetilde{z}}^{(#1)}_{#2} }}
\newcommand{\tavars}{\ensuremath{ \widetilde{\mathsf{Var}} }} 
\newcommand{\ttvars}[1]{\ensuremath{ \widetilde{\mathsf{TVar}}_{#1} }}
\newcommand{\formalpha}{\ensuremath{ \widetilde{\Sigma}} }

\newcommand{\eqformulae}[1]{\mathit{Eq}(#1)}

\newcommand{\letterofrule}{\alpha} 
\newcommand{\formulaoftree}{\Phi} 
\newcommand{\formulaoftreefree}{\Psi} 

\newcommand{\tavarbegin}[1]{\widetilde{b_{#1}}}
\newcommand{\tavarend}[1]{\widetilde{e_{#1}}}

\newcommand{\tavarcomp}[1]{\xi_{#1}}
\newcommand{\tavarinter}[1]{\zeta_{#1}}

\lstdefinelanguage{JavaScript}{
  keywords={typeof, true, false, catch, function, return, null, catch, switch, var, if, in, while, do, od, else, case, break, when, with, assume},
  ndkeywords={class, export, boolean, throw, implements, import, this},
  sensitive=false,
  comment=[l]{//},
  morecomment=[s]{/*}{*/},
  morecomment=[s]{$}{$},
  morestring=[b]',
  morestring=[b]"
}
\lstnewenvironment{code}[1][]{%
    \lstset{%
	    basicstyle=\codesize\ttfamily, 
	    breaklines=false,
	    commentstyle=\color[HTML]{073642}\ttfamily\itshape,
	    identifierstyle=\color{black},
	    inputencoding=latin1,
	    keywordstyle=\color[HTML]{228B22}\bfseries,
	    language=JavaScript,
	    ndkeywordstyle=\color[HTML]{228B22}\bfseries,
	    numbers=left,
	    numbersep=5pt,
	    xleftmargin=30pt,
	    xrightmargin=10pt,
	    frame=Tb,
	    framexleftmargin=15pt,
	    numberstyle=\scriptsize, 
	    prebreak = \raisebox{0ex}[0ex][0ex]{\ensuremath{\hookleftarrow}},
	    showstringspaces=false,
	    stringstyle=\color[rgb]{0.639,0.082,0.082}\ttfamily,
	    upquote=true,
	    mathescape=true,
    	escapebegin=\color[HTML]{073642},
	    tabsize=2,
	    #1
    }
}{}

\lstnewenvironment{examplecode}[1][]{%
    \lstset{%
      basicstyle=\codesize\ttfamily, 
      breaklines=false,
      commentstyle=\color[HTML]{073642}\ttfamily\itshape,
      identifierstyle=\color{black},
      inputencoding=latin1,
      keywordstyle=\color[HTML]{228B22}\bfseries,
      language=JavaScript,
      ndkeywordstyle=\color[HTML]{228B22}\bfseries,
      numbers=left,
      numbersep=5pt,
      xleftmargin=75pt,
      xrightmargin=60pt,
      frame=Tb,
      framexleftmargin=15pt,
      numberstyle=\scriptsize, 
      prebreak = \raisebox{0ex}[0ex][0ex]{\ensuremath{\hookleftarrow}},
      showstringspaces=false,
      stringstyle=\color[rgb]{0.639,0.082,0.082}\ttfamily,
      upquote=true,
      mathescape=true,
      escapebegin=\color[HTML]{073642},
      tabsize=2,
      #1
    }
}{}

\begin{document}

\title{On an Invariance Problem for Parameterized Concurrent Systems}

\author{Marius Bozga \and Lucas Bueri \and Radu Iosif\inst{1}}
\institute{Univ. Grenoble Alpes, CNRS, Grenoble INP, VERIMAG, 38000, France}

\maketitle

\begin{abstract}
  We consider concurrent systems consisting of replicated finite-state
  processes that synchronize via joint interactions in a network with
  user-defined topology. The system is specified using a resource
  logic with a multiplicative connective and inductively defined
  predicates, reminiscent of Separation Logic \cite{Reynolds02}. The
  problem we consider is if a given formula in this logic defines an
  invariant, namely whether any model of the formula, following an
  arbitrary firing sequence of interactions, is transformed into
  another model of the same formula. This property, called \emph{havoc
    invariance}, is quintessential in proving the correctness of
  reconfiguration programs that change the structure of the network at
  runtime. We show that the havoc invariance problem is many-one
  reducible to the entailment problem $\phi \models \psi$, asking if
  any model of $\phi$ is also a model of $\psi$. Although, in general,
  havoc invariance is found to be undecidable, this reduction allows
  to prove that havoc invariance is in \twoexptime, for a general
  fragment of the logic, with a \twoexptime\ entailment problem.
\end{abstract}

\section{Introduction}

The parameterized verification problem asks to decide whether a system
consisting of an arbitrary number of finite-state processes that
communicate via synchronized (joint) actions satisfies a
specification, such as deadlock freedom, mutual exclusion or a
temporal logic property e.g., every request is eventually
answered. The literature in this area has a wealth of decidability and
complexity results (see
\cite{BloemJacobsKhalimovKonnovRubinVeithWidder15} for a survey)
classified according to the communication type (e.g., rendez-vous,
broadcast) and the network topology e.g., rings where every process
interacts with its left/right neighbours, cliques where each two
process may interact, stars with a controller interacting with
unboundedly many workers, etc.

As modern computing systems are dynamically adaptive, recent effort
has been put into designing \emph{reconfigurable systems}, whose
network topologies change at runtime (see
\cite{DBLP:journals/sigact/FoersterS19} for a survey) in order to
address maintenance (e.g., replacement of faulty and obsolete
components by new ones, firmware updates, etc.) and internal traffic
issues (e.g., re-routing to avoid congestion in a datacenter
\cite{DBLP:journals/comsur/Noormohammadpour18}). Unfortunately the
verification of dynamic reconfigurable systems (i.e., proving the
absence of design errors) remains largely unexplored. Consequently,
such systems are prone to bugs that may result in e.g., denial of
services or data corruption\footnote{Google reports on a cascading
cloud failure due to reconfiguration:
\url{https://status.cloud.google.com/incident/appengine/19007}}.

Proving correctness of parameterized reconfigurable networks is
tackled in \cite{AhrensBozgaIosifKatoen21}, where a Hoare-style
program logic is proposed to write proofs of reconfiguration programs
i.e., programs that dynamically add and remove processes and
interactions from the network during runtime. The assertion language
used by these proofs is a logic that describes sets of configurations
defining the network topology and the local states of the
processes. The logic views processes and interactions as resources
that can be joined via a \emph{separating conjunction}, in the spirit
of Separation Logic \cite{Reynolds02}. The separating conjunction
supports \emph{local reasoning}, which is the ability of describing
reconfigurations \emph{only with respect to those components and
interactions that are involved in the mutation}, while disregarding
the rest of the system's configuration. Moreover, the separating
conjunction allows to concisely describe networks of unbounded size,
that share a similar architectural style (e.g., pipelines, rings,
stars, trees) by means of inductively defined predicates.

Due to the interleaving of reconfigurations and interactions between
components, the annotations of the reconfiguration program form a
valid proof under so-called \emph{havoc invariance} assumptions,
stating global properties about the configurations, that remain,
moreover, \emph{unchanged under the ongoing interactions in the
system}. These assumptions are needed to apply the sequential
composition rule that infers a Hoare triple $\hoare{\phi}{P;Q}{\psi}$
from two premisses $\hoare{\phi}{P}{\theta}$ and
$\hoare{\theta}{Q}{\psi}$, where $P$ and $Q$ are reconfiguration
actions that add and/or remove processes and communication
channels. Essentially, because the states of the processes described
by the intermediate assertion $\theta$ might change between the end of
$\mathsf{P}$ and the beginning of $\mathsf{Q}$, this rule is sound
provided that $\theta$ is a havoc invariant formula. 

This paper contributes to the automated generation of reconfiguration
proofs, by a giving a procedure that discharges the havoc invariance
side conditions. The challenge is that a formula of the configuration
logic (that contains inductively defined predicates) describes an
infinite set of configurations of arbitrary sizes. The main result is
that the havoc invariance problem is effectively many-one reducible to
the entailment problem $\phi \models \psi$, that asks if every model
of a formula $\phi$ is a model of another formula $\psi$. Here $\psi$
is the formula whose havoc invariance is being checked and $\phi$
defines the set of configurations $\aconfig'$ obtained from a model
$\aconfig$ of $\psi,$ by executing one interaction from
$\aconfig$. The reduction is polynomial if certain parameters are
bound by a constant (i.e., the arity of the predicates, the size of
interactions and the number of predicate atoms is an inductive rule),
providing a \twoexptime\ upper bound for a fragment of the logic with
a decidable (\twoexptime) entailment problem
\cite[\S6]{BozgaBueriIosif22Arxiv}. Having a polynomial reduction
motivates, moreover, future work on the definition of fragments of
lower (e.g., polynomial) entailment complexity (see e.g.,
\cite{DBLP:conf/concur/CookHOPW11} for a fragment of Separation Logic
with a polynomial entailment problem), that are likely to yield
efficient decision procedures for the havoc invariance problem as
well. In addition, we provide a \twoexptime-hard lower bound for the
havoc invariance problem in this fragment of the logic (i.e., assuming
predicates of unbounded arity) and show that havoc invariance is
undecidable, when unrestricted formul{\ae} are considered as input.
\ifLongVersion \else For space reasons, the technical proofs are given
in Appendix \ref{app:proofs}.  \fi

\noindent
\textbf{Related Work} Specifying parameterized concurrent systems by
inductive definitions is reminiscent of \emph{network grammars}
\cite{ShtadlerGrumberg89,LeMetayer,Hirsch}, that use inductive rules
to describe systems with linear (pipeline, token-ring) architectures
obtained by composition of an unbounded number of processes. In
contrast, we use predicates of unrestricted arities to describe
network topologies that can be, in general, more complex than
trees. Moreover, we write inductive definitions using a resource
logic, suitable also for writing Hoare logic proofs of reconfiguration
programs, based on local reasoning \cite{CalcagnoOHearnYan07}.

Verification of network grammars against safety properties
(reachability of error configurations) requires the synthesis of
\emph{network invariants} \cite{WolperLovinfosse89}, computed by
rather costly fixpoint iterations \cite{LesensHalbwachsRaymond97} or
by abstracting (forgetting the particular values of indices in) the
composition of a small bounded number of instances
\cite{KestenPnueliShaharZuck02}. In previous work, we have developped
an invariant synthesis method based on \emph{structural invariants},
that are synthesized with little computational effort and prove to be
efficient in many practical examples
\cite{DBLP:conf/tacas/BozgaEISW20,DBLP:conf/facs2/BozgaI21}.

The havoc invariance problem considered in this paper is, however,
different from safety checking and has not been addressed before, to
the best of our knowledge. An explaination is that verification of
reconfigurable systems has received fairly scant attention, relying
mostly on runtime verification
\cite{BucchiaroneG08,DormoyKL10,LanoixDK11,DBLP:conf/sac/El-HokayemBS21},
instead of deductive verification, reported in
\cite{AhrensBozgaIosifKatoen21}. In \cite{AhrensBozgaIosifKatoen21} we
addressed havoc invariance with a set of inference rules used to write
proofs manually, whereas the goal of this paper is to discharge such
conditions automatically.

\vspace*{-.5\baselineskip}
\subsection{A Motivating Example}
\label{sec:motivating-example}
\begin{figure}[t!]
  \vspace*{-\baselineskip}
  \begin{center}
    \begin{minipage}{.55\textwidth}
      \hspace*{-5mm}\centerline{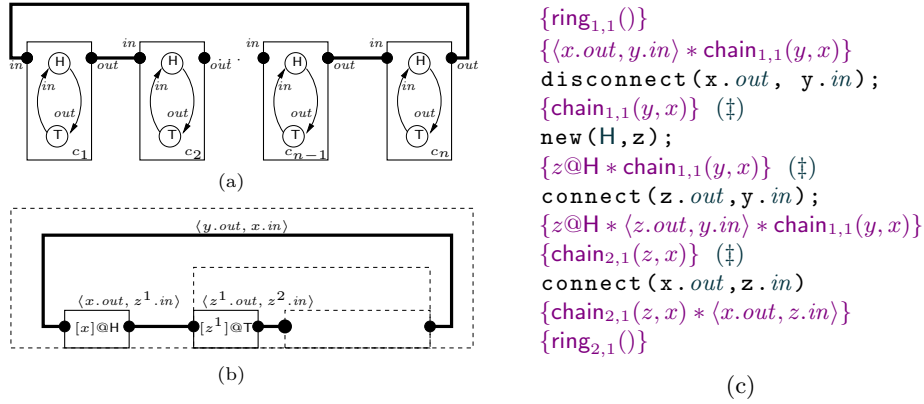}
    \end{minipage}
    \begin{minipage}{.44\textwidth}
      {\small\begin{lstlisting}
$\textcolor{violet}{\set{\ring{1}{1}()}}$        
$\textcolor{violet}{\set{\interac{x}{out}{y}{in} * \chain{1}{1}(y,x)}}$
disconnect(x.$\mathit{out}$, y.$\mathit{in}$);
$\textcolor{violet}{\set{\chain{1}{1}(y,x)}}$ $(\ddagger)$
new($\toknotok$,z);
$\textcolor{violet}{\set{ \compin{z}{\toknotok} * \chain{1}{1}(y,x)}}$ $(\ddagger)$
connect(z.$\mathit{out}$,y.$\mathit{in}$);
$\textcolor{violet}{\set{ \compin{z}{\toknotok} * \interac{z}{out}{y}{in} * \chain{1}{1}(y,x)}}$
$\textcolor{violet}{\set{\chain{2}{1}(z,x)}}$ $(\ddagger)$
connect(x.$\mathit{out}$,z.$\mathit{in}$)
$\textcolor{violet}{\set{\chain{2}{1}(z,x) * \interac{x}{out}{z}{in}}}$
$\textcolor{violet}{\set{\ring{2}{1}()}}$
    \end{lstlisting}}
  \centerline{\footnotesize(c)}
    \end{minipage}
  \end{center}
  \vspace*{-1.5\baselineskip}
  \caption{Inductive Specification and Reconfiguration of a Token Ring}
  \label{fig:ring}
  \vspace*{-.5\baselineskip}
\end{figure}

Consider, for instance, a system consisting of a finite but unbounded
number of processes, called \emph{components} in the following. The
components execute the same machine with states $\toktoken$ and
$\toknotok$, denoting whether the component has a token ($\toktoken$)
or a hole ($\toknotok$). The components are placed in a ring, each
component having exactly one left and one right neighbour, as in
Fig. \ref{fig:ring} (a). A component without a token may receive one,
by executing a transition $\toknotok \arrow{\mathit{in}}{} \toktoken$,
simultaneously with its left neighbour, that executes the transition
$\toktoken \arrow{\mathit{out}}{} \toknotok$ simultaneously, as in
Fig. \ref{fig:ring} (a). Note that there can be more than one token,
moving independently in the system, such that no token overtakes
another token. The configurations of the token ring system are
described by the following inductive rules:
\begin{align*}
\ring{h}{t}() & \leftarrow \exists x \exists y ~.~  \interactwo{x}{out}{y}{in} * \chain{h}{t}(y,x) \\
\chain{h}{t}(x,y) & \leftarrow \exists z.~\compactin{x}{q} * \interac{x}{out}{z}{in} * \chain{h'}{t'}(z,y) \text{, for both } q \in \set{\toknotok,\toktoken} \\
\chain{0}{1}(x,x) & \leftarrow \compactin{x}{\toktoken} \hspace*{8mm}
\chain{1}{0}(x,x) \leftarrow \compactin{x}{\toknotok} \hspace*{8mm}
\chain{0}{0}(x,x) \leftarrow \compact{x} \\
\text{where } h' & \isdef \left\{\begin{array}{ll} \max(h-1,0) & \text{, if } q = \toknotok \\
h & \text{, if } q = \toktoken \end{array}\right. \text{ and }
t' \isdef \left\{\begin{array}{ll} \max(t-1,0) & \text{, if } q = \toktoken \\
t & \text{, if } q = \toknotok \end{array}\right. 
\end{align*}
The predicate $\ring{h}{t}()$ describes a ring with at least $h$ ($t$)
components in state $\toknotok$ ($\toktoken$). The ring consists of an
interaction between the ports $\mathit{out}$ and $\mathit{in}$ of two
components $x$ and $y$, respectively, described by
$\interactwo{x}{out}{y}{in}$ and a separate chain of components
between $x$ and $y$, described by $\chain{h}{t}(y,x)$.  Inductively, a
chain consists of a component $\compactin{x}{q}$ in state $q \in
\set{\toknotok,\toktoken}$, an interaction
$\interactwo{x}{out}{z}{in}$ and a separate $\chain{h'}{t'}(z,y)$,
where $h'$ and $t'$ are the least numbers of components in state
$\toknotok$ and $\toktoken$, respectively, after the removal of the
component $x$. Fig. \ref{fig:ring} (b) depicts the unfolding of the
inductive definition of $\ring{h}{t}()$ with the existentially
quantified variables $z$ from the above rules $\alpha$-renamed to
$z^1, z^2$, etc.  

A \emph{reconfiguration action} is an atomic creation or deletion of a
component or interaction. A \emph{reconfiguration sequence} is a
finite sequence of reconfiguration actions that takes as input a
mapping of program variables to components and executes the actions
from the sequence, in interleaving with the interactions in the
system. For instance, the reconfiguration sequence from
Fig. \ref{fig:ring} (c) takes as input the mapping of $\mathtt{x}$ and
$\mathtt{y}$ to two adjacent components in the token ring, removes the
interaction $\interac{x}{out}{y}{in}$ by executing
$\mathtt{disconnect(x.\mathit{out}, y.\mathit{in})}$ and creates a new
component in state $\toknotok$ (by executing
$\mathtt{new(x,\toknotok)}$) that is connected in between $\mathtt{x}$
and $\mathtt{y}$ via two new interactions created by executing
$\mathtt{connect(z.\mathit{out},y.\mathit{in})}$ and
$\mathtt{connect(x.\mathit{out},z.\mathit{in})}$, respectively.
Fig. \ref{fig:ring} (c) shows a proof (with annotations in curly
braces) of the fact that outcome of the reconfiguration of a ring of
components is a ring whose least number of components in state
$\toknotok$ is increased from one to two. This proof is split into
several subgoals: \begin{compactenum}
\item Entailments required to apply the consequence rule of Hoare
  logic e.g., 
  \(\ring{1}{1}() \models \exists x \exists y ~.~ \interac{x}{out}{y}{in} * \chain{1}{1}(y,x)\).
  The entailment problem has been addressed in
  \cite[\S6]{BozgaBueriIosif22Arxiv}, with the definition of a general
  fragment of the configuration logic, for which the entailment problem
  is decidable in double exponential time.
\item Hoare triples that describe the effect of the atomic
  reconfiguration actions e.g., \(\{\interac{x}{out}{y}{in} *
  \chain{1}{1}(y,x)\} \mathtt{disconnect(x.\mathit{out},
    y.\mathit{in})} \{\chain{1}{1}(y,x)\}\). These are obtained by
  applying the frame rule to the local\footnote{A Hoare triple
  $\hoare{\phi}{P}{\psi}$ is \emph{local} if it mentions only those
  components and interactions added or deleted by $\mathsf{P}$. Local
  specifications are plugged into a global context by the \emph{frame
  rule} that infers $\hoare{\phi*F}{P}{\psi*F}$ from
  $\hoare{\phi}{P}{\psi}$ if the variables modified by $\mathsf{P}$
  are not free in $F$.}  specifications of the atomic actions.
  The local specification of reconfiguration actions and the frame
  rule for local actions are described in
  \cite[\S4.2]{AhrensBozgaIosifKatoen21}.
\item \emph{Havoc invariance} proofs for the annotations marked with
  ($\ddagger$) in Fig. \ref{fig:ring} (c). For instance, the formula
  $\chain{1}{1}(y,x)$ is havoc invariant because the interactions in a
  chain of components will only move tokens to the right without
  creating more or losing any, hence there will be the same number of
  components in state $\toknotok$ ($\toktoken$) no matter which
  interactions are fired.
\end{compactenum}

\section{Definitions}
\label{sec:definitions}

We denote by $\nat$ the set of positive integers, including zero. For
a set $A$, we denote $A^1 \isdef A$, $A^{i+1} \isdef A^i \times A$,
for all $i \geq 0$, where $\times$ denotes the Cartesian product, and
$A^+ \isdef \bigcup_{i\geq1} A^i$. The cardinality of a finite set $A$
is denoted by $\cardof{A}$. By writing $A \finsubseteq B$ we mean that
$A$ is a finite subset of $B$. Given integers $i$ and $j$, we write
$\interv{i}{j}$ for the set $\set{i,i+1,\ldots,j}$, assumed to be
empty if $i>j$. For a function $f : A \rightarrow B$, we denote by
$f[a_i \leftarrow b_i]_{i \in \interv{1}{n}}$ the function that maps
$a_i$ into $b_i$ for each $i \in \interv{1}{n}$ and agrees with $f$
everywhere else. 

\vspace*{-.5\baselineskip}
\subsection{Configurations}
We model a parallel system as a hypergraph, whose vertices are
\emph{components} (i.e., the nodes of the network) and hyperedges are
\emph{interactions} (i.e., describing the way the components
communicate with each other). The components are taken from a
countably infinite set $\universe$, called the \emph{universe}. We
consider that each component executes its own copy of the same
\emph{behavior}, represented as a finite-state machine
$\beh=(\ports,\states,\arrow{}{})$, where $\ports$ is a finite set of
\emph{ports}, $\states$ is a finite set of \emph{states} and
$\arrow{}{} \subseteq \states \times \ports \times \states$ is a
transition relation. Intuitively, each transition $q \arrow{p}{} q'$
of the behavior $\beh$ is triggerred by a visible event, represented
by the port $p$.
The universe $\universe$ and the behavior
$\beh=(\ports,\states,\arrow{}{})$ are considered to be fixed in the
following.

A \emph{configuration} is a snapshot of the system, describing the
topology of the network (i.e., the set of present components and
interactions) together with the local state of each component,
formally defined below:

\begin{definition}\label{def:configuration}
  A \emph{configuration} is a tuple $\aconfig = (\comps,\interacs, \statemap)$, where: 
  \begin{compactitem}
  \item $\comps \finsubseteq \universe$ is a finite set of
    \emph{components}, that are present in the configuration,
  \item $\interacs \finsubseteq (\universe\times\ports)^+$ is a finite
    set of \emph{interactions}, where each interaction is a sequence
    $(c_i, p_i)_{i \in \interv{1}{n}} \in (\universe \times \ports)^n$
    that binds together the ports $p_1, \ldots, p_n$ of the pairwise
    distinct components $c_1, \ldots, c_n$, respectively. The ordered
    sequence of ports $(p_1, \ldots, p_n)$ is called an
    \emph{interaction type} and we denote by $\ports^+$ the set of
    interaction types.
  \item $\statemap : \universe \rightarrow \states$ is a \emph{state
    map} associating each (possibly absent) component, a state of the
    behavior $\beh$, such that the set $\set{c \in \universe \mid
      \statemap(c) = q}$ is infinite, for each $q \in \states$.
  \end{compactitem}
  We denote by $\configset$ the set of configurations.
\end{definition}
The last condition requires that there is an infinite pool of
components in each state $q \in \states$; since $\universe$ is
infinite and $\states$ is finite, this condition is feasible.

\begin{example}\label{ex:configuration}
  The configurations of the system from Fig. \ref{fig:ring} (a) are
  $(\{c_1, \ldots, c_n\}, \{(c_i,\mathit{out}, \\ c_{(i \mod n) +
    1},\mathit{in}) \mid i \in \interv{1}{n}\}, \statemap)$, where
  $\statemap:\universe\rightarrow\set{\toknotok,\toktoken}$ is a state
  map. The ring topology is given by components $\{c_1, \ldots, c_n\}$
  and interactions $\{(c_i,\mathit{out}, c_{(i \mod n) +
    1},\mathit{in}) \mid i \in \interv{1}{n}\}$.  \hfill$\blacksquare$
\end{example}

Note that Def. \ref{def:configuration} allows configurations with
interactions that involve absent components i.e., not from the set
$\comps$ of present components in the given configuration. The
following definition distinguishes such configurations:

\begin{definition}\label{def:tightness}
  A configuration $\aconfig = (\comps,\interacs,\statemap)$ is said to
  be \emph{tight} if and only if for any interaction $(c_i, p_i)_{i
    \in \interv{1}{n}} \in \interacs$ we have $\set{c_i \mid i \in
    \interv{1}{n}} \subseteq \comps$ and \emph{loose} otherwise.
\end{definition}
For instance, every configuration of the system from
Fig. \ref{fig:ring} (a) is tight and becomes loose if a component is
deleted.

\vspace*{-.5\baselineskip}
\subsection{Configuration Logic}

Let $\vars$ and $\preds$ be countably infinite sets of \emph{variables} and \emph{predicates}, respectively.
For each predicate $\apred \in \preds$, we denote its arity by $\arityof{\apred}$.
The formul{\ae} of the \emph{Configuration Logic} (\cl) are described inductively by the following syntax:
\vspace*{-.25\baselineskip}
\[\phi := \emp \mid \compact{x} \mid \interacn{x_1}{p_1}{x_n}{p_n} \mid \compin{x}{q}
\mid x=y \mid x\neq y \mid \apred(x_1, \ldots, x_{\arityof{\apred}})
\mid \phi * \phi \mid \exists x ~.~ \phi\] where $x, y, x_1, \ldots
\in \vars$, $q \in \states$ and $\apred \in \preds$.  A formula
$\compact{x}$, $\interacn{x_1}{p_1}{x_n}{p_n}$, $\compin{x}{q}$ and
$\apred(x_1, \ldots, x_{\arityof{\apred}})$ is called a
\emph{component}, \emph{interaction}, \emph{state} and
\emph{predicate} atom, respectively. We use the shorthand
$\compactin{x}{q} \isdef \compact{x} * \compin{x}{q}$.  Intuitively, a
formula $\compactin{x}{q} * \compactin{y}{q'} *
\interactwo{x}{out}{y}{in} * \interactwo{x}{in}{y}{out}$ describes a
configuration consisting of two distinct components, denoted by the
values of $x$ and $y$, in states $q$ and $q'$, respectively, and two
interactions binding the $\mathit{out}$ port of one to the
$\mathit{in}$ port of the other component.

A formula with no occurrences of predicate atoms (resp. existential
quantifiers) is called \emph{predicate-free}
(resp. \emph{quantifier-free}). A \qpf\ formula is both predicate- and
quantifier-free. A variable is \emph{free} if it does not occur in the
scope of a quantifier and $\fv{\phi}$ is the set of free variables of
$\phi$. 
A \emph{substitution} $\phi[x_i/y_i]_{i\in\interv{1}{n}}$ replaces
simultaneously every free occurrence of $x_i$ by $y_i$ in $\phi$, for
all $i \in \interv{1}{n}$.  The size of a formula $\phi$ is the total
number of occurrences of symbols needed to write it down, denoted by
$\sizeof{\phi}$.

The only connective of the logic is the \emph{separating conjunction}
$*$. Intuitively, $\phi_1 * \phi_2$ means that $\phi_1$ and $\phi_2$
hold separately, on disjoint parts of the same configuration. Its
formal meaning is coined by the following definition of
\emph{composition} of configurations:

\begin{definition}\label{def:composition}
  The composition of two configurations $\aconfig_i = (\comps_i,
  \interacs_i, \statemap)$, for $i = 1,2$, such that $\comps_1 \cap
  \comps_2 = \emptyset$ and $\interacs_1 \cap \interacs_2 =
  \emptyset$, is defined as $\aconfig_1 \comp \aconfig_2 \isdef
  (\comps_1 \cup \comps_2, \interacs_1 \cup \interacs_2, \statemap)$.
  The composition $\aconfig_1 \comp \aconfig_2$ is undefined if
  $\comps_1 \cap \comps_2 \neq \emptyset$ or $\interacs_1 \cap
  \interacs_2 \neq \emptyset$.
\end{definition}

\begin{example}\label{ex:composition}
  Let $\aconfig_i = (\set{c_i}, \set{(c_i, \mathit{out}, c_{3-i},
    \mathit{in})}, \statemap)$ be configurations, for $i = 1,2$. Then
  $\aconfig_1 \comp \aconfig_2 = (\set{c_1, c_2}, \set{(c_1,
    \mathit{out}, c_2, \mathit{in}), (c_2, \mathit{out}, c_1,
    \mathit{in})}, \statemap)$. \hfill$\blacksquare$
\end{example}

The meaning of the predicates is given by a set of inductive
definitions:

\begin{definition}\label{def:sid}
  A \emph{set of inductive definitions (SID)} $\asid$ consists of
  \emph{rules} of the form $\apred(x_1, \ldots, x_{\arityof{\apred}})
  \leftarrow \phi$, where $x_1, \ldots, x_{\arityof{\apred}}$ are
  pairwise distinct variables, called \emph{parameters}, such that
  $\fv{\phi} \subseteq \set{x_1, \ldots, x_{\arityof{\apred}}}$.  We
  say that the rule $\apred(x_1, \ldots, x_{\arityof{\apred}})
  \leftarrow \phi$ \emph{defines} $\apred$ and denote by
  $\defn{\asid}{\apred}$ the set of rules from $\asid$ that define
  $\apred$ and by $\defnof{\asid} \isdef \set{\apred \mid
    \defn{\asid}{\apred} \neq \emptyset}$ the set of predicates
  defined by $\asid$.
\end{definition}
Note that having distinct parameters in a rule is without loss of
generality, as e.g., a rule $\apred(x_1, x_1) \leftarrow \phi$ can be
equivalently written as $\apred(x_1, x_2) \leftarrow x_1 = x_2 *
\phi$. As a convention, we shall always use the names $x_1, \ldots,
x_{\arityof{\apred}}$ for the parameters of a rule that defines
$\apred$. An example of a SID is given in
\S\ref{sec:motivating-example}.

The size of a SID is $\sizeof{\asid} \isdef \sum_{\apred(x_1, \ldots,
  x_{\arityof{\apred}}) \leftarrow \phi \in \asid} \sizeof{\phi} +
\arityof{\apred} + 1$. Other parameters, relevant for complexity
evaluation, are the maximal \begin{inparaenum}[(1)]
\item arity $\maxarityof{\asid} \isdef \max\{\arityof{\apred} \mid
  \apred(x_1, \ldots, x_{\arityof{\apred}}) \leftarrow \phi \in
  \asid\}$ of a defined predicate,
\item size of an interaction type $\maxinterof{\asid}
  \isdef \max\{n \mid \interacn{y_1}{p_1}{y_n}{p_n} \text{ occurs in }
  \asid\}$, and
\item number of predicate atoms $\maxpredsof{\asid} \isdef \max\{h
  \mid \apred(x_1, \ldots, x_{\arityof{\apred}}) \leftarrow \exists
  y_1 \ldots \exists y_m ~.~ \phi * \Asterisk_{\ell=1}^h
  \bpred_\ell(\vec{z}_\ell), \phi \text{ is a \qpf\ formula}\}$. 
\end{inparaenum}

The semantics of \cl\ formul{\ae} is defined by a satisfaction
relation $\aconfig \models^\store_\asid \phi$ between configurations
and formul{\ae}.  This relation is parameterized by a \emph{store}
$\store : \vars \rightarrow \universe$ mapping the free variables of a
formula into components from the universe (possibly absent from
$\aconfig$) and an SID $\asid$. The definition of the satisfaction
relation is by induction on the structure of formul{\ae}, where
$\aconfig = (\comps, \interacs, \statemap)$ is a configuration
(Def. \ref{def:configuration}):
\[\begin{array}{rclcl}
\aconfig & \models^\store_\asid & \emp & \iff & \comps = \emptyset \text{ and } \interacs = \emptyset \\
\aconfig & \models^\store_\asid & \compact{x} & \iff & \comps = \set{\store(x)} \text{ and } \interacs = \emptyset \\
\aconfig & \models^\store_\asid & \interacn{x_1}{p_1}{x_n}{p_n} & \iff & \comps = \emptyset \text{ and } \interacs = \set{(\nu(x_1), p_1, \ldots, \nu(x_n), p_n)} \\
\aconfig & \models^\store_\asid & \compin{x}{q} & \iff & \aconfig \models^\store_\asid \emp \text{ and } \statemap(\store(x)) = q \\
\aconfig & \models^\store_\asid & x \sim y & \iff & \aconfig \models^\store_\asid \emp \text{ and } \store(x)\sim\store(y) \text{, for all } \sim \in \set{=,\neq} \\
\aconfig & \models^\store_\asid & \apred(y_1, \ldots, y_{\arityof{\apred}}) & \iff & \aconfig \models^\store_\asid \phi[x_1/y_1, \ldots, x_{\arityof{\apred}}/y_{\arityof{\apred}}] \text{, for some rule } \\
&&&& \apred(x_1, \ldots, x_{\arityof{\apred}}) \leftarrow \phi \text{ from } \asid \\
\aconfig & \models^\store_\asid & \phi_1 * \phi_2 & \iff & \text{there exist } \aconfig_1 \text{ and } \aconfig_2 \text{, such that } \aconfig = \aconfig_1 \comp \aconfig_2 \text{ and } \\
&&&& \aconfig_i \models^\store_\asid \phi_i \text{, for all } i = 1,2 \\
\aconfig & \models^\store_\asid & \exists x ~.~ \phi & \iff & \aconfig \models^{\store[x \leftarrow c]}_\asid \phi \text{, for some } c \in \universe
\end{array}\]
If $\aconfig \models^\store_\asid \phi$, we say that the pair
$(\aconfig,\store)$ is a $\asid$-model of $\phi$.
If $\phi$ is a predicate-free formula, the satisfaction relation does
not depend on the SID, written $\aconfig \models^\store \phi$. A
formula $\phi$ is \emph{satisfiable} if and only if it has a model. A
formula $\phi$ $\asid$-\emph{entails} a formula $\psi$, written $\phi
\models_\asid \psi$, if and only if any $\asid$-model of $\phi$ is a
$\asid$-model of $\psi$. Two formul{\ae} are
$\asid$-\emph{equivalent}, written $\phi \equiv_\asid \psi$ if and
only if $\phi \models_\asid \psi$ and $\psi \models_\asid \phi$. A
formula $\phi$ is $\asid$-\emph{tight} if $\aconfig$ is tight
(Def. \ref{def:tightness}), for any $\asid$-model $(\aconfig,\store)$
of $\phi$. We omit mentioning $\asid$ whenever it is clear from the
context or not needed.

\vspace*{-.5\baselineskip}
\subsection{The Havoc Invariance Problem}

This paper is concerned with the \emph{havoc invariance} problem i.e.,
the problem of deciding whether the set of models of a given
\cl\ formula is closed under the execution of a sequence of
interactions. The execution of an interaction $(c_i, p_i)_{i \in
  \interv{1}{n}}$ synchronizes transitions labeled by the ports $p_1,
\ldots, p_n$ from the behaviors (i.e., replicas of the state machine
$\beh$) of $c_1, \ldots, c_n$, respectively. This joint execution of
several transitions in different components of the system is formally
described by the \emph{step} relation below:

\begin{definition}\label{def:havoc}
  The \emph{step} relation $\step{} \subseteq \configset \times
  (\universe \times \ports)^+ \times \configset$ is defined as:
  \vspace*{-.7\baselineskip}
  \[\begin{array}{l}
  (\comps,\interacs,\statemap) \step{(c_i,p_i)_{i \in \interv{1}{n}}}
  (\comps,\interacs,\statemap') \text{ if and only if } {(c_i,p_i)}_{i \in \interv{1}{n}}
  \in \interacs \text{ and } \statemap' = \statemap[c_i \leftarrow
    q'_i] _{i \in \interv{1}{n}} \\
  \text{where } \statemap(c_i) = q_i \text{ and }
  q_i \arrow{p_i}{} q_i' \text{ is a transition of } \beh \text{, for all } i \in
  \interv{1}{n}
  \end{array}\]
  The \emph{havoc} relation $\Havoc$ is the reflexive and transitive
  closure of the relation $\Step \subseteq \configset^2$:
  $(\comps,\interacs,\statemap) \Step (\comps,\interacs,\statemap')$
  if and only if $(\comps,\interacs,\statemap) \step{(c_i,p_i)_{i \in
      \interv{1}{n}}} (\comps,\interacs,\statemap')$, for some
  interaction ${(c_i,p_i)}_{i \in \interv{1}{n}} \in \interacs$.
\end{definition}

\begin{example}\label{ex:havoc}
  Let $\aconfig_i = (\set{c_1, c_2, c_3}, \set{(c_i, \mathit{out},
    c_{i \mod 3 + 1}, \mathit{in}) \mid i \in \interv{1}{3}},
  \statemap_i)$, for $i \in \interv{1}{3}$ be configurations, where
  $\statemap_1(c_1) = \statemap_1(c_2) = \toknotok$, $\statemap_1(c_3)
  = \toktoken$, $\statemap_2(c_1) = \toktoken$, $\statemap_2(c_2) =
  \statemap_2(c_3) = \toknotok$, $\statemap_3(c_1) = \statemap_3(c_3)
  = \toknotok$, $\statemap_3(c_2) = \toktoken$. Then we have
  $\aconfig_i \Havoc \aconfig_j$, for all $i,j \in
  \interv{1}{3}$. \hfill$\blacksquare$
\end{example}

Two interactions $(c_1, p_1, \ldots, c_n, p_n)$ and $(c_{i_1},
p_{i_1}, \ldots, c_{i_n}, p_{i_n})$ such that $\set{i_1, \ldots, i_n}
= \interv{1}{n}$, are equivalent from the point of view of the step
relation, since the set of executed transitions is the same;
nevertheless, we chose to distinguish them in the following, for
reasons of simplicity. Note, moreover, that the havoc relation does
not change the component or the interaction set of a configuration,
only its state map.

\begin{definition}\label{def:havoc-invariance}
  Given an SID $\asid$ and a predicate $\apred$, the problem
  $\hinv{\asid}{\apred}$ asks whether for all $\aconfig,\aconfig' \in
  \configset$ and each store $\store$, such that $\aconfig
  \models_\asid^\store \apred(x_1,\dots,x_{\arityof{\apred}})$ and
  $\aconfig \Havoc \aconfig'$, it is the case that $\aconfig'
  \models_\asid^\store \apred(x_1,\dots,x_{\arityof{\apred}})$?
\end{definition}

\begin{example}\label{ex:havoc-invariance}
  Consider a model $\aconfig = (\{c_1, \ldots, c_n\},
  \{(c_i,\mathit{out}, c_{(i \mod n) + 1},\mathit{in}) \mid i \in
  \interv{1}{n}\}, \statemap)$ of the formula $\ring{1}{1}()$ i.e.,
  having the property that $\statemap(c_i) = \toknotok$ and
  $\statemap(c_j) = \toktoken$ for at least two indices $i \neq j \in
  \interv{1}{n}$, where the SID that defines $\ring{1}{1}()$ is given
  in \S\ref{sec:motivating-example}. Similar to Example
  \ref{ex:havoc}, in any configuration $\aconfig' = (\{c_1, \ldots,
  c_n\}, \{(c_i,\mathit{out}, c_{(i \mod n) + 1},\mathit{in}) \mid i
  \in \interv{1}{n}\}, \statemap')$ such that $\aconfig \Havoc
  \aconfig'$, we have $\statemap'(c_k) = \toknotok$ and
  $\statemap'(c_\ell) = \toktoken$, for some $k \neq \ell \in
  \interv{1}{n}$, hence $\aconfig'$ is a model of $\ring{1}{1}()$,
  meaning that $\ring{1}{1}()$ is havoc invariant. Examples of
  formul{\ae} that are not havoc invariant include e.g., 
  $\compactin{x}{\toktoken} * \interac{x}{\outp}{y}{\inp} *
  \compactin{y}{\toknotok}$. \hfill$\blacksquare$
\end{example}

Without loss of generality, we consider the havoc invariance problem
only for single predicate atoms. This is because, for any formula
$\phi$, such that $\fv{\phi} = \set{x_1,\ldots,x_n}$, one may consider
a fresh predicate symbol (i.e., not in the SID) $\apred_\phi$ and add
the rule $\apred_\phi(x_1, \ldots, x_n) \leftarrow \phi$ to the
SID. Then $\phi$ is havoc invariant if and only if $\apred_\phi(x_1,
\ldots, x_n)$ is havoc invariant.



\section{From Havoc Invariance to Entailment}
\label{sec:reduction}

We describe a many-one reduction of the havoc invariance
(Def. \ref{def:havoc-invariance}) to the entailment problem, following
three steps. Given an instace $\hinv{\asid}{\apred}$ of the havoc
invariance problem, the SID $\asid$ is first translated into a tree
automaton recognizing trees labeled with predicate-free formul{\ae},
that symbolically encode the set of $\asid$-models of the predicate
atom $\apred(x_1, \ldots, x_{\arityof{\apred}})$. Second, we define a
structure-preserving tree transducer that simulates the effect of
executing exactly one interaction from such a model. Third, we compute
the image of the language recognized by the first tree automaton via
the transducer, as a second tree automaton, which is translated back
into another SID $\overline{\asid}$ defining one or more predicates
$\overline{\apred}_1, \ldots, \overline{\apred}_p$, among
other. Finally, we prove that $\hinv{\asid}{\apred}$ has a positive
answer if and only if each of the entailments
$\set{\overline{\apred}_i(x_1, \ldots, x_{\arityof{\apred}})
  \models_{\asid \cup \overline{\asid}} \apred(x_1, \ldots,
  x_{\arityof{\apred}})}_{i=1}^p$ produced by the reduction, hold.

For the sake of self-containment, we recall below the definitions of
trees, tree automata and (structure-preserving) tree transducers. Let
$(\Sigma,\arityof{})$ be a ranked alphabet, where each symbol $\alpha \in
\Sigma$ has an associated arity $\arityof{\alpha}\geq0$. A \emph{tree} over
$\Sigma$ is a finite partial function $\tree : \nat^* \finmap \Sigma$,
whose domain $\dom{\tree} \finsubseteq \nat^*$ is both
\emph{prefix-closed} i.e., $u\in\dom{t}$, for all $u,v\in\nat^*$, such
that $u \cdot v\in\dom{\tree}$, and \emph{complete} i.e., $\set{n \in
  \nat \mid u \cdot n\in\dom{t}}=\interv{1}{\arityof{\tree(u)}}$, for
all $u \in \dom{\tree}$. Given $u \in \dom{\tree}$, the \emph{subtree}
of $\tree$ rooted at $u$ is the tree $\subtree{\tree}{u}$, such that
$\dom{\subtree{t}{u}} \isdef \set{w \mid u\cdot w \in \dom{\tree}}$
and $\subtree{\tree}{u}(w) \isdef \tree(u \cdot w)$. We denote by
$\treesof{\Sigma}$ the set of trees over a ranked alphabet $\Sigma$. 

A \emph{tree automaton} (TA) is a tuple
$\auth=(\Sigma,\tastates,\tafinstates,\tatrans)$, where $\Sigma$ is a
ranked alphabet, $\tastates$ is a finite set of states,
$\tafinstates\subseteq\tastates$ is a set of final states and
$\tatrans$ is a set of transitions $\alpha(s_1,\ldots,s_{\arityof{a}})
\arrow{}{} s$; when $\arityof{\alpha}=0$, we write $\alpha \arrow{}{}
s$ instead of $\alpha() \arrow{}{} s$. A \emph{run} of $\auth$ over a
tree $\tree$ is a function $\run : \dom{\tree} \rightarrow \tastates$,
such that, for all $u \in \dom{\tree}$, we have $\run(u)=s$ if
$(\tree(u))(\pi(u \cdot 1), \ldots, \pi(u \cdot \arityof{\tree(u)}))
\arrow{}{} s \in\tatrans$. Given a state $q \in \tastates$, a run
$\run$ of is $q$-\emph{accepting} if and only if $\run(\epsilon)=q$,
in which case $\auth$ is said to $q$-accept $\tree$. We denote by
$\langof{q}{\auth}$ the set of trees $q$-accepted by $\auth$ and let
$\lang{\auth} \isdef \bigcup_{q \in \tafinstates}
\langof{q}{\auth}$. A language $L$ is \emph{recognizable} if and only
if there exists a TA $\auth$, such that $L = \lang{\auth}$.

A \emph{tree transducer} (TT) is a tree automaton over an alphabet of
pairs $\trans = (\Sigma^2,\tastates,\tafinstates,\tatrans)$, such that
$\arityof{\alpha} = \arityof{\beta} = n$, for each transition
$(\alpha,\beta)(s_1,\ldots,s_n) \arrow{}{} s \in
\tatrans$. Intuitively, a transition of the transducer reads a symbol
$\alpha$ from the input tree and writes another symbol $\beta$ to the
output tree, at the same position. Cleary, any tree $\tree : \nat^*
\finmap \Sigma^2$ with labels from the set of pairs
$\set{(\alpha,\beta) \in \Sigma^2 \mid
  \arityof{\alpha}=\arityof{\beta}}$ can be viewed as a pair of trees
$(\tree_1,\tree_2)$ over $\Sigma$, such that
$\dom{\tree_1}=\dom{\tree_2}=\dom{\tree}$. In order to define the
image of a tree language via a transducer, we
define \begin{inparaenum}[(i)]
\item \emph{projection} $\proj{i}{L} \isdef \set{\tree_i \mid
 (\tree_1,\tree_2) \in L}$, for all $i=1,2$, where $L \subseteq
  \treesof{\Sigma^2}$, and
\item \emph{cylindrification} $\cyl{i}{L} \isdef
  \set{(\tree_1,\tree_2) \mid \tree_i \in L}$, for all $i=1,2$, where
  $L \subseteq \treesof{\Sigma}$.
\end{inparaenum}
The \emph{image} of a language $L \subseteq \treesof{\Sigma}$ via a
transducer $\trans$ is the language $\trans(L) \isdef
\proj{2}{\left(\cyl{1}{L} \cap \lang{\trans}\right)}$. It is manifest
that $\trans(L)$ is recognizable whenever $L$ is recognizable.

\vspace*{-.5\baselineskip}
\subsection{From SID to Tree Automata and Back}

We define a two-way connection between SIDs and TAs, as
follows: \begin{compactenum}
\item Given a finite SID $\asid$ we define a TA $\auth_\asid$, whose
  states $q_\apred$ are named after the predicates $\apred$ that occur
  in $\asid$ and whose alphabet consists of the predicate-free
  formul{\ae} from the rules of $\asid$, with variables mapped to
  canonical names, together with a tuple of arities, needed for later
  bookkeeping. Each tree $\tree \in \langof{q_\apred}{\auth_\asid}$
  defines a unique predicate-free formula $\formulaoftree(\tree)$,
  such that the $\asid$-models of a predicate atom $\apred(x_1,
  \ldots, x_{\arityof{\apred}})$ are exactly the models of some
  $\formulaoftree(\tree)$, for $\tree \in
  \langof{q_\apred}{\auth_\asid}$.
\item Conversely, given a TA $\auth$ over an alphabet of formul{\ae}
  annotated with arities, the tuple of arities associated with each
  alphabet symbol allows to define a SID $\asid_\auth$, whose
  predicates $\apred_q$ are named after the states $q$ of the TA, such
  that the models of the formul{\ae} $\formulaoftree(\tree)$, such
  that $\tree \in \langof{q}{\auth}$ are exactly the
  $\asid_\auth$-models of the predicate atom $\apred_q(x_1, \ldots,
  x_{\arityof{\apred_q}})$.
\end{compactenum}
Let us fix a countably infinite set of variables \(\tavars \isdef
\set{\tavarin{i} \mid i \geq 1} \cup \set{\tavarout{\ell}{i} \mid
  i,\ell \geq 1}\), with the understanding that $\tavarin{i}$ are
canonical names for the variables from the left-hand side and
$\tavarout{\ell}{i}$ are canonical names for the variables occurring
in the $\ell$-th predicate atom from the right-hand side of a rule. An
\emph{alphabet symbol} $\alpha = \tuple{\psi,a_0,\ldots,a_h}$ consists
of a predicate-free formula $\psi$ and a tuple of positive integers
$a_0, \ldots, a_h \in \nat$, such that $\fv{\psi} = \set{\tavarin{i}
  \mid i \in \interv{1}{a_0}} \cup 
\set{\tavarout{\ell}{i} \mid \ell \in \interv{1}{h},~ i \in
  \interv{1}{a_\ell}}$. We take the arity of such a symbol to be
$\arityof{\alpha} \isdef h$ and denote by $\formalpha$ the (infinite)
set of alphabet symbols. Trees labeled with symbols from $\formalpha$
define predicate-free \emph{characteristic formul{\ae}}, as follows:

\begin{definition}\label{def:charform}
  Given a tree $\tree \in \treesof{\formalpha}$, where
  $\tree(\epsilon) = \tuple{\exists y_1 \dots \exists y_{m}~.~ \phi, a_0, \ldots, a_h}$ 
  with $\phi$ a \qpf\ formula, and a node $u \in
  \nat^*$, we define the \qpf\ \emph{characteristic formula}: 
  $$\formulaoftreefree^u(\tree) \isdef ~\phi \substitution{\tavarin{j}}{x_j^u}{j\in\interv{1}{a_0}}
   \substitution{\tavarout{\ell}{j}}{x_j^{u\cdot \ell}}{
       \ell\in\interv{1}{h},
       j\in\interv{1}{a_\ell}
   } \substitution{y_j}{y_j^u}{j\in\interv{1}{m}} *
    \Asterisk_{\hspace*{-1mm}\ell \in \interv{1}{h}} ~\formulaoftreefree^{u\cdot\ell}(\subtree{\tree}{\ell})$$  
    Assuming that $\tree(v) = \tuple{\exists y_1 \dots \exists
    y_{m^v}~.~ \phi^v, a_0^v, \ldots, a_h^v}$, for all $v\in\dom{t}$, we
  consider also the predicate-free formula $\formulaoftree^u(\tree) =
  (\exists x_j^{u\cdot v})_{v\in\dom{t}\setminus \set{\epsilon},~ j\in\interv{1}{a_0^v}}
  (\exists y_j^{u\cdot v})_{v\in\dom{t},~ j\in\interv{1}{m^v}} 
  ~.~\formulaoftreefree^u(\tree)$.
  
\end{definition}
\ifLongVersion 
Note that every variable $x$ 
in $\formulaoftreefree^\epsilon(\tree)$ is superscripted with the node $u
\in \dom{\tree}$, such that $x$ occurs (free or quantified) in $\tree(u)$.
More precisely, we have $\fv{\formulaoftreefree^u(\tree)} =
\set{x_j^{u\cdot v} \mid v\in\dom{t}, j\in\interv{1}{a_0^v}} \cup
\set{y_j^{u\cdot v} \mid v\in\dom{t}, j\in\interv{1}{m^v}}$ and
$\fv{\formulaoftree^u(\tree)} = \set{x_j^u \mid
  j\in\interv{1}{a_0^\epsilon}}$. \fi

\begin{figure}[t!]
  \centerline{\scalebox{0.85}{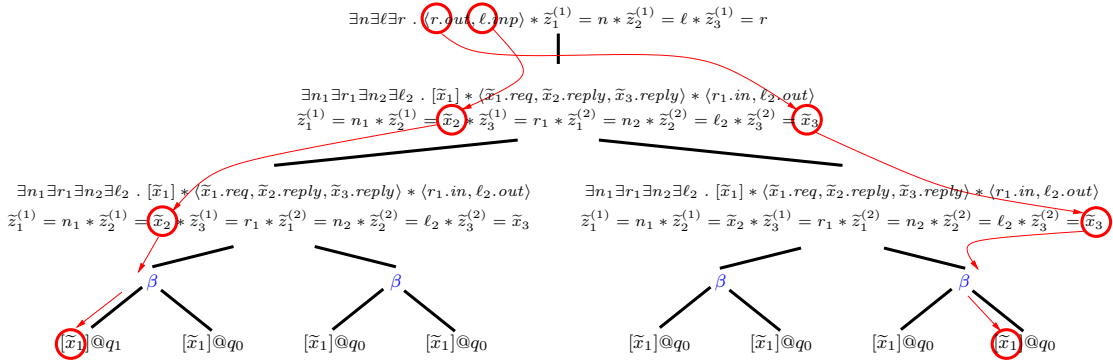}}
  \caption{Tree Labeled with Formul{\ae} Encoding a System from Example \ref{ex:tll}}
  \label{fig:tree}
  \vspace*{-.5\baselineskip}
\end{figure}

\begin{example}\label{ex:tll}  
  \ifLongVersion
  \begin{figure}[t!]
    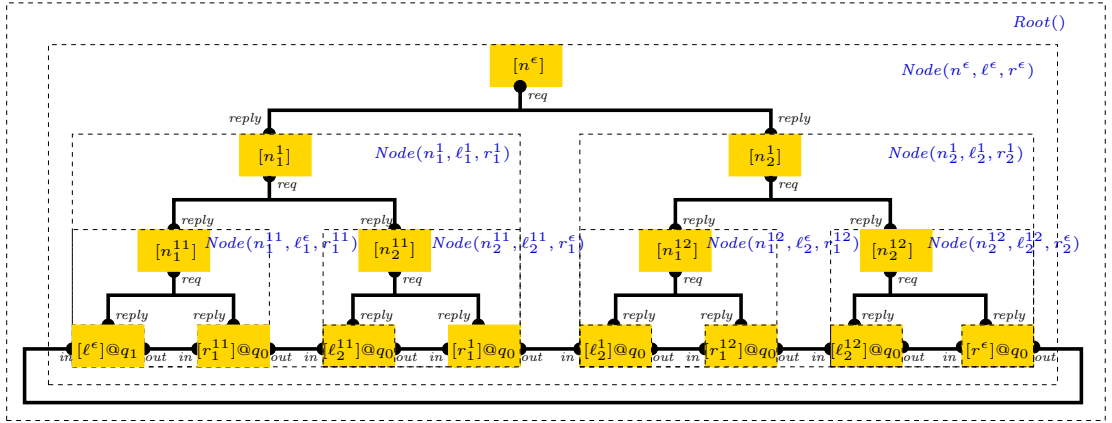
    \caption{Inductive Specification of a Tree with Leaves Linked in a Ring}
    \label{fig:tll}
  \end{figure}
  Fig. \ref{fig:tll} shows \else We consider \fi a system whose
  components form a tree, in which each parent sends a request
  ($\req$) to and receives replies ($\reply$) from both its
  children. In addition, the leaves of the tree form a ring, with the
  $\outp$ port of each leaf connected to the $\inp$ port of its right
  neighbour. The system is described by the following inductive
  definitions:
  \vspace*{-.5\baselineskip}
  \begin{align}
    \treeroot() \leftarrow & \exists n \exists \ell \exists r ~.~
    \tuple{r.\outp,\ell.\inp} * \treenode(n,\ell,r) \label{rule:root} \\
    \treenode(n,\ell,r) \leftarrow & \exists n_1 \exists r_1 \exists n_2 \exists \ell_2 ~.~
    \compact{n} * \tuple{n.\req,n_1.\reply,n_2.\reply} * \tuple{r_1.\inp, \ell_2.\outp} ~* \nonumber \\
    & \hspace*{4.5cm} \treenode(n_1,\ell,r_1) * \treenode(n_2,\ell_2,r) \label{rule:node} \\
    \treenode(n,\ell,r) \leftarrow & \compactin{n}{q_0} \hspace*{1.5cm} \treenode(n,\ell,r) \leftarrow \compactin{n}{q_1} \label{rule:leaf}
  \end{align}
  Fig. \ref{fig:tree} shows a tree $\tree \in \treesof{\formalpha}$
  describing an instance of the system, where $\formalpha =
  \set{\alpha,\beta,\gamma_0,\gamma_1}$:
  \vspace*{-.5\baselineskip}
  \begin{align*}
    \hspace*{-4mm} \alpha \isdef & \tuple{\exists n \exists \ell \exists r ~.~
      \tuple{r.\outp,\ell.\inp} * \tavarout{1}{1} = n * \tavarout{1}{2} = \ell * \tavarout{1}{3} = r, 0, 3} \\
    \hspace*{-4mm} \beta \isdef & \langle\exists n_1 \exists r_1 \exists n_2 \exists \ell_2 ~.~
    \compact{n} * \tuple{n.\req,\ell.\reply,r.\reply} * \tuple{r_1.\inp, \ell_2.\outp} ~* \\
    & \tavarout{1}{1} = n_1 * \tavarout{1}{2} = \ell * \tavarout{1}{3} = r_1 *
    \tavarout{2}{1} = n_2 * \tavarout{2}{2} = \ell * \tavarout{2}{3} = r, 3,3,3 \rangle \\
    \hspace*{-4mm} \gamma_0 \isdef & \tuple{\compactin{\tavarin{1}}{q_0},3} \hspace*{3cm}
    \gamma_1 \isdef \tuple{\compactin{\tavarin{1}}{q_1},3}
  \end{align*}
  For simplicity, Fig. \ref{fig:tree} shows only the formul{\ae}, not
  the arity lists of the alphabet symbols. \hfill$\blacksquare$
\end{example}

The models of the characteristic formula
$\formulaoftreefree^\epsilon(\tree)$ of a tree $\tree \in
\treesof{\formalpha}$ define walks in the tree that correspond to
chains of equalities between variables. Formally, a \emph{walk} in
$\tree$ is a sequence of nodes $u_1, \ldots, u_n \in \dom{\tree}$,
such that $u_i$ is either the parent or a child of $u_{i+1}$, for all
$i \in \interv{1}{n-1}$. Note that a walk can visit the same node of
the tree several times. In particular, if the characteristic formula
$\formulaoftreefree^\epsilon(\tree)$ is tight (i.e., has only tight
models in the sense of Def. \ref{def:tightness}) there exist equality
walks between the node containing an interaction atom and the nodes
where these variables are instantiated by component atoms. For
instance, walks between the root containing
$\interac{r}{\outp}{\ell}{\inp}$ and the left- and right-most leafs,
labeled with component atoms that associate elements of $\universe$ to
the variables $\ell$ and $r$ are shown in Fig. \ref{fig:tree}.

\begin{lemmaE}\label{lemma:walks}
  Let $\tree \in \treesof{\formalpha}$ be a tree, such that
  $\formulaoftreefree^\epsilon(\tree)$ is tight, $(\aconfig,\store)$
  be a model of $\formulaoftreefree^\epsilon(\tree)$ and $y^v$, $z^w$
  be two variables that occur in a component and interaction atom of
  $\formulaoftreefree^\epsilon(\tree)$, respectively. Then
  $\store(y^v) = \store(z^w)$ if and only if there exists a walk $u_1,
  \dots, u_n$ in $\tree$ and variables $y^v = x^{u_1}_{i_1}, \ldots,
  x^{u_n}_{i_n} = z^w$, such that either $x^{u_j}_{i_j}$ and
  $x^{u_{j+1}}_{i_{j+1}}$ are the same variable, or the equality atom
  $x^{u_j}_{i_j} = x^{u_{j+1}}_{i_{j+1}}$ occurs in
  $\formulaoftreefree^\epsilon(\tree)$, for all $j \in
  \interv{1}{n-1}$.
\end{lemmaE}
\begin{proofE}
  By Def.~\ref{def:charform}, each equality atom within
  $\formulaoftreefree^\epsilon(\tree)$ is of the form $x^u = y^v$,
  such that either $u=v$, $u$ is the parent of $v$, or $u$ is a child
  of $v$ in $\tree$. Suppose, for a contradiction, that there is no
  walk $u_1, \dots, u_n \in \dom{\tree}$ with the above property. Then
  we build a loose model $(\aconfig',\store')$ of
  $\formulaoftreefree^\epsilon(\tree)$. Let  $\aconfig=(\comps,\interacs,\statemap)$, $c = \store(x^u)$ and $\overline{c}
  \in \universe \setminus (\comps \cup \set{c_i \mid
    (c_i,p_i)_{i\in\interv{1}{n}} \in \interacs})$ be a component not
  occurring in $\aconfig$. We define $(\aconfig',\store')$ as
  follows: \begin{compactitem}
  \item $\aconfig' = (\comps,\interacs',\statemap)$, where $\interacs'
    = \{(c_i,p_i)_{i\in\interv{1}{n}} \in \interacs \mid c \not\in
      \set{c_1, \ldots, c_n}\} \cup \{(c'_i,p_i)_{i\in\interv{1}{n}}
      \mid (c_i,p_i)_{i\in\interv{1}{n}} \in \interacs,~ c \in
      \set{c_1, \ldots, c_n},~ c'_i = c_i \text{ if } c_i \neq c
      \text{ and } c'_i = \overline{c} \text{ otherwise}\}$.
  \item for each variable $x \in \vars$, we have $\nu'(x) =
    \overline{c}$ if $\formulaoftreefree^\epsilon(\tree) \models z^w =
    x$ and $\nu'(x) = \nu(x)$, otherwise.
  \end{compactitem} 
  Clearly, $\aconfig'$ is loose, because $\overline{c}$ occurs in an
  interaction from $\interacs'$ but $\overline{c} \not\in \comps$. We
  conclude by showing $(\aconfig',\nu') \models
  \formulaoftreefree^\epsilon(\tree)$, by induction on the structure
  of $\tree$. However, this contradicts with the assumption that
  $\formulaoftreefree^\epsilon(\tree)$ is tight, which concludes the
  proof.
\end{proofE}


Let $\asid$ be a fixed and finite SID in the following. We build a TA
$\auth_\asid$ that recognizes the $\asid$-models of each predicate
atom defined by $\asid$, in the sense of Lemma \ref{lemma:sid-ta}
below.

\begin{definition}\label{def:sid-ta}
  We associate each rule \(r : \ruleform\in\asid\), where $\phi$ is
  a \qpf\ formula, with the alphabet symbol:
  \vspace*{-.5\baselineskip}
  \[\letterofrule_r \isdef
  \Big\langle\exists y_1 \ldots \exists y_m ~.~ \Big(\phi * \Asterisk_{      
        \ell \in \interv{1}{h},~ 
        i \in \interv{1}{\arityof{\bpred_\ell}}
    }
    ~\tavarout{\ell}{i} = z^\ell_i \Big)
    \substitution{x_j}{\tavarin{j}}{j \in
      \interv{1}{\arityof{\apred}}}, \arityof{\apred},
    \arityof{\bpred_1}, \ldots, \arityof{\bpred_h}\Big\rangle\] Let
    $\auth_\asid \isdef (\talpha_\asid,\tastates_\asid,
    \tatrans_\asid)$ be a TA, where $\talpha_\asid \isdef
    \set{\letterofrule_r \mid r \in \asid}$, $\tastates_\asid \isdef
    \set{\tastatepred{\apred} \mid \apred\in \defnof{\asid}}$ and
    $\tatrans_\asid \isdef \set{\letterofrule_r
      (\tastatepred{\bpred_1}, \dots, \tastatepred{\bpred_h})
      \rightarrow \tastatepred{\apred} \mid r \in \asid}$.
\end{definition}



\begin{example}[contd. from Example \ref{ex:tll}]\label{ex:tll-ta}
  The TA corresponding to the SID in Example \ref{ex:tll} is
  $\auth_\asid = (\formalpha,\tastates_\asid,\tatrans_\asid)$, where
  $\formalpha = \set{\alpha,\beta,\gamma_0,\gamma_1}$,
  $\tastates_\asid = \set{\tastatepred{\treeroot},
    \tastatepred{\treenode}}$ and $\tatrans_\asid =
  \set{\alpha(\tastatepred{\treenode}) \rightarrow
    \tastatepred{\treeroot},
    \beta(\tastatepred{\treenode},\tastatepred{\treenode}) \rightarrow
    \tastatepred{\treenode}, \gamma_0 \rightarrow
    \tastatepred{\treenode}, \gamma_1 \rightarrow
    \tastatepred{\treenode}}$. \hfill$\blacksquare$
\end{example}

The following lemma proves that the predicate-free formul{\ae}
corresponding (in the sense of Def. \ref{def:charform}) to the trees
recognized by $\auth_\asid$ in a state $q_\apred$ define the
$\asid$-models of the predicate atom $\apred(x_1, \ldots,
x_{\arityof{\apred}})$:


\begin{lemmaE}\label{lemma:sid-ta}
For any predicate $\apred \in \defnof{\asid}$, configuration
$\aconfig$, store $\store$ and node $u \in \nat^*$, we have $\aconfig
\models^\store_\asid \apred(x_1^u, \dots, x_{\arityof{\apred}}^u)$ if
and only if $\aconfig \models^\store \formulaoftree^u(\tree)$, for
some tree $\tree\in \langof{q_\apred}{\auth_\asid}$.
\end{lemmaE}
\begin{proofE}
  For a store $\store$, we denote by $\store[x_i/y_i]_{i \in
    \interv{1}{n}}$ the store that maps $x_i$ to $\store(y_i)$ and
  agrees with $\store$ everywhere else.

  \vspace*{\baselineskip}\noindent``$\Rightarrow$'' By induction on
  the definition of the satisfaction relation
  $\aconfig\models^\store_\asid \apred(x_1^u, \dots,
  x_{\arityof{\apred}}^u)$. Assume that there exists a rule
  $r\in\asid$ of the form $\ruleform$, where $\phi$ is a
  \qpf\ formula, and a store $\store'$ such that $\store'(x_j) =
  \store(x_j^u)$ for all $j \in \interv{1}{\arity}$, where $\aconfig =
  \aconfig_0 \comp \ldots \comp \aconfig_h$ is such that $\aconfig_0
  \models^{\store'} \phi$ and $\aconfig_\ell \models^{\store'}_\asid
  \bpred_\ell (z^\ell_1, \ldots, z^\ell_{\arityof{\bpred_\ell}})$, for
  all $\ell \in \interv{1}{h}$. Hence, we obtain \(\aconfig_\ell
  \models^{\store'_\ell}_\asid \bpred_\ell (x^{u \cdot \ell}_1,
  \ldots, x^{u \cdot \ell}_{\arityof{\bpred_\ell}})\), where
  $\store'_\ell \isdef \store'[x_i^{u \cdot \ell}/z_i^\ell]_{i \in
    \interv{1}{\arityof{\bpred_\ell}}}$, for all $\ell \in
  \interv{1}{h}$. By the induction hypothesis, there exist trees
  $\tree_\ell \in \langof{q_{\bpred_\ell}}{\auth_\asid}$, such that
  $\aconfig_\ell \models^{\store'_\ell}_\asid \formulaoftree^{u \cdot
    \ell}(\tree_\ell)$, for each $\ell \in \interv{1}{h}$. We define
  the tree $\tree$ as: \begin{compactitem}
  \item $\dom{\tree} \isdef \set{\epsilon} \cup
    \bigcup_{\ell\in\interv{1}{h}} \ell \cdot \dom{\tree_\ell}$,
  \item $\tree(\epsilon) \isdef \tuple{ \psi, \arityof{\apred},
    \arityof{\bpred_1}, \ldots, \arityof{\bpred_h}}$, where $\psi
    \isdef \exists y_1 \ldots \exists y_m ~.~ \Big(\phi * \Asterisk_{
      \hspace*{-2mm}
      \begin{array}{l}
        \scriptstyle{\ell \in \interv{1}{h}} \\[-2mm]
        \scriptstyle{i \in \interv{1}{\arityof{\bpred_\ell}}}
    \end{array}}
    \hspace*{-4mm}\tavarout{\ell}{i} = z^\ell_i \Big)
    \substitution{x_i}{\tavarin{i}}{i \in \interv{1}{\arityof{\apred}}}$, 
  \item $\subtree{\tree}{\ell} \isdef \tree_\ell$, for all $\ell \in
    \interv{1}{h}$.
  \end{compactitem}
  By the definition of $\auth_\asid$ and $\tree$, we obtain $\tree \in
  \langof{q_\apred}{\auth_\asid}$ and we are left with proving that
  $\aconfig \models^\store_\asid \formulaoftree^u(\tree)$. To this
  end, we define the store $\store''$, such that: \begin{compactitem}
  \item $\store''(x^u_j) = \store(x^u_j)$, for all $j \in \interv{1}{\arityof{\apred}}$, 
  \item $\store''(y^u_k) = \store'(y_k)$, for all $k \in \interv{1}{m}$, 
  \item $\store''(x^{u \cdot \ell}_i) = \store'_\ell(x^{u \cdot
    \ell}_i)$, for all $i \in \interv{1}{\bpred_\ell}$ and $\ell \in
    \interv{1}{h}$.
  \end{compactitem}
  By the definition of $\store''$, we obtain:
  \[\aconfig \models^{\store''}_\asid \psi\substitution{\tavarin{i}}{x_i^u}{i\in\interv{1}{\arityof{\apred}}}
   \substitution{\tavarout{\ell}{i}}{x_i^{u\cdot \ell}}{\hspace*{-1mm} \begin{array}{l}
       \scriptstyle{i\in\interv{1}{\arityof{\bpred_\ell}}} \\[-2mm]
       \scriptstyle{\ell\in\interv{1}{h}}
   \end{array}} \hspace*{-2mm} \substitution{y_k}{y_k^u}{k\in\interv{1}{m}} *
    \Asterisk_{\ell \in \interv{1}{h}}
    \formulaoftree^{u\cdot\ell}(\subtree{\tree}{\ell})\] leading to
    $\aconfig \models^\store_\asid \formulaoftree^u(\tree)$, because
    $\store''$ agrees with $\store$ over $x^u_1, \ldots,
    x^u_{\arityof{\apred}}$.

    \vspace*{\baselineskip} \noindent''$\Leftarrow$'' Let $\tree\in
    \langof{q_\apred}{\auth_\asid}$ be a tree such that $\aconfig
    \models^\store \formulaoftree^u(\tree)$. We proceed by induction
    on the structure of $\tree$. Let be $\letterofrule_r \isdef
    \tree(\epsilon)$ the label of the root of $\tree$, and
    $\subtree{\tree}{1}, \ldots, \subtree{\tree}{h}$ be the subtrees
    rooted in the children of the root of $\tree$ ($h$ can be equal to
    $0$ if $\tree$ consists of a leaf). Since $\tree \in
    \langof{q_\apred}{\auth_\asid}$, there exists a rule $r \in \asid$
    and a transition $\letterofrule_r(\tastatepred{\bpred_1}, \dots,
    \tastatepred{\bpred_h}) \to \tastatepred{\apred}$ in
    $\auth_\asid$, where $r$ is of the form $\ruleform$, for a
    \qpf\ formula $\phi$, such that $\subtree{\tree}{\ell} \in
    \langof{q_{\bpred_\ell}}{\auth_\asid}$, for all $\ell \in
    \interv{1}{h}$. By Def. \ref{def:sid-ta}, we have:
    \[\letterofrule_r = \Tuple{\exists
    y_1 \ldots \exists y_m ~.~ \Big(\phi * \Asterisk_{
      \hspace*{-2mm}
      \begin{array}{l}
        \scriptstyle{\ell \in \interv{1}{h}} \\[-2mm]
        \scriptstyle{i \in \interv{1}{\arityof{\bpred_\ell}}}
    \end{array}}
    \hspace*{-4mm}\tavarout{\ell}{i} = z^\ell_i \Big)
    \substitution{x_i}{\tavarin{i}}{i \in
      \interv{1}{\arityof{\apred}}}, \arityof{\apred},
    \arityof{\bpred_1}, \ldots, \arityof{\bpred_h}}\] Since $\aconfig
    \models^\store \formulaoftree^u(\tree)$, by
    Def. \ref{def:charform}, there exists a store $\store'$ that
    agrees with $\store$ over $x^u_1, \ldots, x^u_{\arityof{\apred}}$
    and configurations $\aconfig = \aconfig_0 \comp \ldots \comp
    \aconfig_h$, such that: \begin{compactitem}
    \item $\aconfig_0 \models^{\store'} \Big(\phi * \Asterisk_{
      \hspace*{-2mm}
      \begin{array}{l}
        \scriptstyle{\ell \in \interv{1}{h}} \\[-2mm]
        \scriptstyle{i \in \interv{1}{\arityof{\bpred_\ell}}}
    \end{array}}
      \hspace*{-4mm}x_i^{u\cdot \ell} = z^\ell_i \Big) \substitution{x_i}{x_i^u}{i \in \interv{1}{\arityof{\apred}}}
      \substitution{y_k}{y_k^u}{k\in\interv{1}{m}}$, 
    \item $\aconfig_\ell \models^{\store'}
      \formulaoftree^{u\cdot\ell}(\subtree{\tree}{\ell})$, for all
      $\ell \in \interv{1}{h}$.
    \end{compactitem}
    Since $\subtree{\tree}{\ell} \in
    \langof{q_{\bpred_\ell}}{\auth_\asid}$, by the inductive
    hypothesis we obtain $\aconfig_\ell \models^{\store'}_\asid
    \bpred_\ell(x^{u \cdot \ell}_1, \ldots, x^{u \cdot
      \ell}_{\arityof{\bpred_\ell}})$, for all $\ell \in
    \interv{1}{h}$. Let us define the store:
    \[\store'' \isdef \store'[z^\ell_i / x^{u \cdot \ell}_i]_{\hspace*{-2mm}\begin{array}{l}
        \scriptstyle{\ell \in \interv{1}{h}} \\[-2mm]
        \scriptstyle{i \in \interv{1}{\arityof{\bpred_\ell}}}
    \end{array}} \hspace*{-4mm} [y_k/y^u_k]_{k \in \interv{1}{m}}\]
    We obtain $\aconfig \models_\asid^{\store''} \phi *
    \Asterisk_{\ell \in \interv{1}{h}} \bpred_\ell(z^{\ell}_1, \ldots,
    z^{\ell}_{\arityof{\bpred_\ell}})$, hence $\aconfig
    \models_\asid^\store \exists y_1 \ldots \exists y_m ~.~ \phi *
    \Asterisk_{\ell \in \interv{1}{h}} \bpred_\ell(z^{\ell}_1, \ldots,
    z^{\ell}_{\arityof{\bpred_\ell}})$, leading to $\aconfig
    \models^\store_\asid \apred(x^u_1, \ldots,
    x^u_{\arityof{\apred}})$.    
\end{proofE}

Conversely, given a tree automaton $\auth =
(\talpha,\tastates,\tatrans)$, we construct a SID $\asid_\auth$ that
defines the models of the predicate-free formul{\ae} corresponding
(Def. \ref{def:charform}) to the trees recognized by $\auth$ (Lemma
\ref{lemma:ta-sid}). We assume that the alphabet $\talpha$ consists of
symbols $\tuple{\psi,a_0,\ldots,a_h}$ of arity $h$, where $\psi$ is a
predicate-free formula with free variables $\fv{\psi} =
\set{\tavarin{i} \mid i \in \interv{1}{a_0}} \cup
\set{\tavarout{\ell}{i} \mid \ell \in \interv{1}{h},~ i \in
  \interv{1}{a_\ell}}$ and that the transitions of the TA meet the
requirement:

\begin{definition}\label{def:sid-compatible}
  A TA $\auth$ is \emph{SID-compatible} iff for any transitions
  $\tuple{\psi,a_0,\ldots,a_h}(q_1,\ldots,q_h) \arrow{}{} q_0$ and
  $\tuple{\psi',a'_0,\ldots,a'_h}(q'_1,\ldots,q'_h) \arrow{}{} q'_0$
  of $\auth$, we have $q_i = q'_i$ only if $a_i = a'_i$, for all $i
  \in \interv{0}{h}$.
\end{definition}
Let us fix a SID-compatible TA $\auth = (\talpha, \tastates,
\tatrans)$ for the rest of this section.

\begin{definition}\label{def:ta-sid}
  The SID $\asid_\auth$ has a rule:
  \vspace*{-.5\baselineskip}
  \begin{align*}
    \hspace*{-5mm} \apred_{q_0}(x_1, \ldots, x_{a_0}) \leftarrow &
    \exists y^1_1 \ldots \exists y^h_{a_h} ~.~ 
    \phi[\tavarin{i}/x_i]_{i \in \interv{1}{a_0}} [\tavarout{\ell}{i}/y^\ell_i]_{      
        \ell \in \interv{1}{h},~ 
        i \in \interv{1}{a_\ell}
        }
     ~* \Asterisk_{\ell\in\interv{1}{h}}
    \apred_{q_\ell}(y^\ell_1, \ldots, y^\ell_{a_\ell})
  \end{align*}
  for each transition $\tuple{\phi, a_0, \ldots, a_h}(q_1, \ldots,
  q_h) \arrow{}{} q_0$ of $\auth$ and those rules only. 
\end{definition}

The following lemma states that $\asid_\auth$ defines the set of
models of the characteristic formul{\ae} (Def. \ref{def:charform}) of
the trees recognized by $\auth$.

\begin{lemmaE}\label{lemma:ta-sid}
  For any state $q \in \tastates$, configuration $\aconfig$, store
  $\store$ and node $u \in \nat^*$, we have $\aconfig
  \models^\store_{\asid_\auth} \apred_q(x_1^u, \dots,
  x_{\arityof{\apred_q}}^u)$ if and only if $\aconfig \models^\store
  \formulaoftree^u(\tree)$, for some tree $\tree\in \langof{q}{\auth}$. 
\end{lemmaE}
\begin{proofE}
  Let $\overline{\auth}$ denote the automaton $\auth_{\asid_\auth}$
  and $\overline{q}$ denote the state $\tastatepred{\apred_q}$ of
  $\overline{\auth}$ (Def. \ref{def:sid-ta}). By Lemma
  \ref{lemma:sid-ta}, we have $\aconfig \models^\store_{\asid_\auth}
  \apred_q(x_1^u, \dots, x_{\arityof{\apred_q}}^u)$ if and only if
  $\aconfig \models^\store \formulaoftree^u(\overline{\tree})$, for
  some tree $\overline{\tree} \in
  \langof{\overline{q}}{\overline{\auth}}$. It is sufficient to prove
  that for each tree $\overline{\tree} \in
  \langof{\overline{q}}{\overline{\auth}}$ there exists a tree $\tree
  \in \langof{q}{\auth}$ such that
  $\formulaoftree^u(\overline{\tree})$ is equivalent to
  $\formulaoftree^u(\tree)$, and viceversa. The last point is a
  consequence of the one-to-one mapping between the transitions of
  $\auth$ and those of $\overline{\auth}$:
  \[\begin{array}{l}
  \tuple{\phi, a_0, \ldots, a_h}(q_1, \ldots, q_h) \rightarrow q_0 \in \tatrans \\
  \iff \text{ by Def. \ref{def:ta-sid}} \\
  \apred_{q_0}(x_1, \ldots, x_{a_0}) \leftarrow \exists y^1_1 \ldots \exists y^1_{a_1} \ldots
    \exists y^h_1 \ldots \exists y^h_{a_h} ~.~ \phi[\tavarin{i}/x_i]_{i \in \interv{1}{a_0}} [\tavarout{\ell}{i}/y^\ell_i]_{
      \hspace*{-1mm} \begin{array}{l}
        \scriptstyle{\ell \in \interv{1}{h}} \\[-2mm]
        \scriptstyle{i \in \interv{1}{a_\ell}}
    \end{array}}
    \hspace*{-3mm} * \Asterisk_{\ell\in\interv{1}{h}}
    \apred_{q_\ell}(y^\ell_1, \ldots, y^\ell_{a_\ell}) \in \asid_\auth \\
    \iff \text{ by Def. \ref{def:sid-ta} and quantifier elimination} \\
    \tuple{\phi, a_0, \ldots, a_h}(\overline{q}_1, \ldots, \overline{q}_h)
    \rightarrow \overline{q}_0 \in \tatrans_{\asid_\auth}
  \end{array}\]
\end{proofE}


\vspace*{-.5\baselineskip}
\subsection{Encoding Havoc Steps by Tree Transducers}

The purpose of this section is the definition of a transducer that
simulates one havoc step. Before giving its definition, we note that
the havoc invariance problem can be equivalently defined by
considering the transformation induced by a single havoc step, instead
of an arbitrary sequence of steps. The following lemma can be taken as
an equivalent definition:

\begin{lemmaE}\label{lemma:havoc-step}
  $\hinv{\asid}{\apred}$ has a positive answer if and only if, for all
  $\aconfig,\aconfig' \in \configset$ and each store $\store$, such
  that $\aconfig \models_\asid^\store
  \apred(x_1,\dots,x_{\arityof{\apred}})$ and $\aconfig \Step
  \aconfig'$, it is the case that $\aconfig' \models_\asid^\store
  \apred(x_1,\dots,x_{\arityof{\apred}})$.
\end{lemmaE}
\begin{proofE}
  Let $\sinv{\asid}{\apred}$ be the decision problem from the
  statement of the Lemma and prove that $\sinv{\asid}{\apred} \iff
  \hinv{\asid}{\apred}$. ``$\Rightarrow$'' Let $\aconfig, \aconfig'
  \in \configset$ and $\store$, such that $\aconfig
  \models_\asid^\store \apred(x_1, \ldots, x_{\arityof{\apred}})$ and
  $\aconfig \Havoc \aconfig'$. Then there exist configurations
  $\aconfig = \aconfig_0, \ldots, \aconfig_n = \aconfig'$, such that
  $\aconfig_i \Step \aconfig_{i+1}$, for all $i \in \interv{0}{n-1}$.
  Proving $\aconfig' \models_\asid^\store \apred(x_1, \ldots,
  x_{\arityof{\apred}})$ goes by induction over $n\geq0$. For the base
  case $n=0$, we have $\aconfig=\aconfig'$ and there is nothing to
  prove. For the inductive step $n>0$, we have $\aconfig_{n-1}
  \models_\asid^\store \apred(x_1,\dots,x_{\arityof{\apred}})$ by the
  inductive hypothesis. Then $\aconfig_{n-1} \Step \aconfig'$ and
  $\aconfig' \models_\asid^{\store}
  \apred(x_1,\dots,x_{\arityof{\apred}})$ follows, by the hypothesis
  $\sinv{\asid}{\apred}$. ``$\Leftarrow$'' Trivial, because
  $\Step\ \subseteq\ \Havoc$.
\end{proofE}

We fix a SID $\asid$ for the rest of this section and recall the
existence of a fixed finite-state behavior
$\beh=(\ports,\states,\arrow{}{})$ with ports $\ports$, states
$\states$ and transitions $q \arrow{p}{} q' \in \states \times \ports
\times \states$. We define a transducer $\trans_\tau$ parameterized by
a given interaction type $\tau = (p_1, \ldots, p_n) \in \ports^+$.
The havoc step transducer is the automata-theoretic union of the typed
transducers over the set of interaction types that occurs in $\asid$.

Given a tree $\tree \in \treesof{\formalpha}$, an interaction-typed
transducer $\trans_\tau$ \begin{inparaenum}[(1)]
\item guesses an interaction atom $\interacn{z_1}{p_1}{z_n}{p_n}$ that
  occurs in some label of $\tree$, 
\item tracks the equality walks (Lemma \ref{lemma:walks}) between each
  variable $z_i$ and the component atom $\compactin{x_i}{q_i}$ that
  defines the store value of $z_i$ and its current state, and 
\item replaces each state component atom $\compactin{x_i}{q_i}$ by
  $\compactin{x_i}{q'_i}$, where $q_i \arrow{p_i}{} q'_i$ is a
  transition from $\beh$, for each $i \in \interv{1}{n}$.
\end{inparaenum}
The output of the transducer is a tree $\tree' \in
\treesof{\formalpha}$, that symbolically encodes the effect of
executing some interaction of type $\tau$ over $\tree$. The main
challenge in defining $\trans_\tau$ is that the equality walks between
an interaction atom $\interacn{z_1}{p_1}{z_n}{p_n}$ and the component
atoms instatiating the variables $z_1, \ldots, z_n$ may visit a tree
node more than once. To capture this, the transducer will guess at
once the equalities summarizing the different fragments of the walk
that lie in the currently processed subtree of $\tree$. Accordingly,
the states of $\trans_\tau$ are conjunctions of equalities, with
special variables $\tavarbegin{i}$ (resp. $\tavarend{i}$) indicating
whether a component (resp. interaction) atom has already been
encountered in the current subtree, intuitively marking the
\emph{beginning} (resp. \emph{end}) of the walk.





For an interaction type $\tau = (p_1, \ldots, p_n)$, let
\(\ttvars{\tau} \isdef \set{\tavarin{i} \mid i \in
  \interv{1}{\maxarityof{\asid}}} \cup \set{\tavarbegin{i},
  \tavarend{i} \mid i\in \interv{1}{n}} \) and
let $\eqformulae{\ttvars{\tau}}$ be the set of separating conjunctions
of equality atoms i.e., $\varphi \isdef \Asterisk_{\!\!i \in I} ~x_i =
y_i$, such that $\fv{\varphi} \subseteq \ttvars{\tau}$. Note that
$\exists x ~.~ \varphi$, for $\varphi \in \eqformulae{\ttvars{\tau}}$, is
equivalent to a formula from $\eqformulae{\ttvars{\tau}}$ obtained by
eliminating the quantifier: either $x$ occurs in an atom $x = y$ for a
variable $y$ distinct from $x$ then $(\exists x ~.~ \varphi) \equiv
\varphi[x/y]$, or $x\not\in\fv{\varphi}$, in which case $(\exists x
~.~ \varphi) \equiv \varphi$.

\begin{definition}\label{def:trans}
The transducer $\trans_{\intertype} \isdef (\talpha_\asid^2,
\tastates_{\intertype}, \tafinstates_{\intertype},
\tatrans_{\intertype})$, where $\tau = (p_1, \ldots, p_n)$, is as follows: \begin{compactitem}
\item $\tastates_{\intertype} = \set{ \varphi \in \eqformulae{\ttvars{\tau}}
  \mid \varphi \not\models (\tavarbegin{i} = \tavarbegin{j}),~
  \varphi \not\models (\tavarend{i} = \tavarend{j}),
  \varphi \not\models (\tavarbegin{i} = \tavarend{j}) \text{, for any } i\neq j}$, 
\item $\tafinstates_\intertype = \set{\varphi \in
  \tastates_\intertype \mid \varphi
  \models \Asterisk_{i\in\interv{1}{n}}(\tavarbegin{i} = \tavarend{i})}$, and
\item $\tatrans_\intertype$ contains transitions of the form
  $(\alpha,\alpha')(\varphi_1, \dots, \varphi_h) \arrow{}{\trans} \varphi$
  where: \begin{compactitem}
  \item $\alpha = (\exists y_1 \ldots \exists y_m ~.~ \psi, a_0,
    \dots, a_h)$ and $\alpha' = (\exists y_1 \ldots \exists y_m ~.~
    \psi', a_0, \dots, a_h)$, where $\psi$ and $\psi'$ are
    \qpf\ formul{\ae} such that $\fv{\psi} = \fv{\psi'} \subseteq
    \ttvars{\tau} \cup \set{y_1, \ldots, y_m}$,
  \item there exists a set $I = \{i_1,\dots, i_r \}
    \subseteq \interv{1}{n}$, variables $\xi_1, \ldots, \xi_r \in
    \fv{\psi}$ and transitions $q_{1} \arrow{p_{i_1}}{} q_{1}'$,
    $\dots$, $q_{r} \arrow{p_{i_r}}{} q_{r}'$ in $\beh$, such that
    $\psi = (\Asterisk_{k\in\interv{1}{r}} \compactin{\xi_k}{q_{k}}) *
    \eta$ and $\psi' = (\Asterisk_{k\in\interv{1}{r}}
    \compactin{\xi_k}{q_{k}'}) * \eta$, for some \qpf\ formula $\eta$,
  \item there exists a set $J \in \set{\emptyset,\interv{1}{n}}$, such
    that $\psi$ contains an interaction atom
    $\interacn{\tavarinter{1}}{p_1}{\tavarinter{n}}{p_n}$ if $J = \interv{1}{n}$,
  \item the sets $I$ and $\set{i \in \interv{1}{n} \mid \tavarbegin{i}
    \in \fv{\varphi_\ell}}_{\ell \in \interv{1}{h}}$ are pairwise
    disjoint,
  \item at most one of the sets $J$, $\set{i \in \interv{1}{n} \mid
    \tavarend{i} \in \fv{\varphi_\ell}}_{\ell \in \interv{1}{h}}$ is
    not empty,
  \item $\varphi$ is the result of eliminating the quantifiers from
    the separating conjunction of equalities:
    \[\hspace*{5mm}\exists \tavarout{1}{1} \ldots \exists \tavarout{h}{a_h}
    \exists y_1 \ldots \exists y_m ~.~
    \Asterisk_{\!\!\ell\in\interv{1}{h}}\varphi_\ell[\tavarin{j}/\tavarout{\ell}{j}]_{j
      \in\interv{1}{a_\ell}} ~*~ \Asterisk_{\!k\in\interv{1}{r}}
    \tavarbegin{i_k} = \tavarcomp{k} ~*~ \Asterisk_{\!\ell\in J}\tavarend{\ell} =
   \tavarinter{\ell} ~*~ \psi_{eq}\] where $\psi_{eq}$ is the separating conjunction
    of the equality atoms from $\psi$.
  \end{compactitem}
\end{compactitem}
\end{definition}

\begin{figure}[t!]
  \vspace*{-\baselineskip}
  \centerline{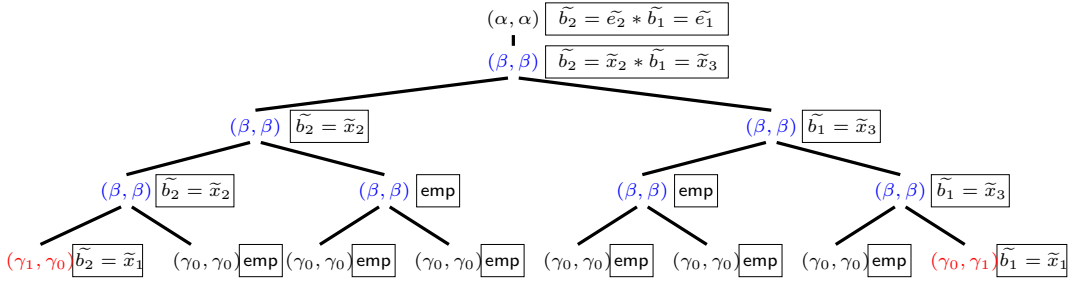}
  \caption{Tree Transducer for the Interactions of Type
    $(\outp,\inp)$ in the System from Fig. \ref{fig:tree}}
  \label{fig:tt}
  \vspace*{-.5\baselineskip}
\end{figure}

\vspace*{-.5\baselineskip}
\begin{example}\label{ex:tt}(contd. from Examples \ref{ex:tll} and \ref{ex:tll-ta})
  Fig. \ref{fig:tt} shows a run of the transducer
  $\trans_{(\outp,\inp)}$, that describes the symbolic execution of
  the interaction corresponding to the
  $\interac{r}{\outp}{\ell}{\inp}$ interaction atom from the root of
  the tree in Fig. \ref{fig:tree}. The states of the transducer are
  separating conjunctions of equality atoms, enclosed within square
  boxes. The transducer replaces the component atoms $\gamma_1 =
  \tuple{\compactin{\tavarin{1}}{q_1},3}$ with $\gamma_0 =
  \tuple{\compactin{\tavarin{1}}{q_0},3}$ (resp. $\gamma_0$ with
  $\gamma_1$) in the left-most (resp. right-most) leaf of the
  tree. \hfill$\blacksquare$
\end{example}

Let $L \subseteq \treesof{\formalpha}$ be an arbitrary language. The
following lemmas prove that the transducer $\trans_\tau$ from
Def. \ref{def:trans} correctly simulates a havoc step produced by an
interaction of type $\tau$.

\begin{lemmaE}\label{lemma:semtransright}
  For each tree $\tree \in L$, such that
  $\formulaoftree^\epsilon(\tree)$ is tight, configurations $\aconfig
  = (\comps,\interacs,\statemap), \aconfig' \in \configset$ and store
  $\store$, such that $\aconfig \models^\store
  \formulaoftree^\epsilon(\tree)$ and $\aconfig \step{(c_i,p_i)_{i \in
      \interv{1}{n}}} \aconfig'$, for some $c_1, \ldots, c_n \in
  \comps$ and $(c_i,p_i)_{i \in \interv{1}{n}} \in \interacs$, there
  exists a tree $\tree' \in \trans_{(p_1,\ldots,p_n)}(L)$, such that
  $\aconfig' \models^\store \formulaoftree^\epsilon(\tree')$.
\end{lemmaE}
\begin{proofE}
  Let $\aconfig \isdef (\comps,\interacs,\statemap)$ and
$\tree(u) \isdef \tuple{\phi^u,a_0^u,\ldots,a_h^u}$, for each
$u \in \dom{\tree}$. Because
$\aconfig \models^\store \formulaoftree^\epsilon(\tree)$, by
Def. \ref{def:charform}, there exists a store $\store'$ that agrees
with $\store$ over $\fv{\formulaoftree^\epsilon(\tree)}$, such that
$\aconfig \models^{\store'} \formulaoftreefree^\epsilon(\tree)$. The
proof is split into the following steps:

\noindent(1) Since there exists an interaction $(c_i,
p_i)_{i \in \interv{1}{n}}$, such that
$\aconfig \step{(c_i,p_i)_{i \in \interv{1}{n}}} \aconfig'$, we define
the tree $\tree'$ based on this interaction. By Def. \ref{def:havoc},
it must be the case that $c_i \in \comps$, for all
$i \in \interv{1}{n}$, $(c_i,p_i)_{i \in \interv{1}{n}} \in \interacs$
and $\aconfig' = (\comps,\interacs,\statemap')$, where
$\statemap'=\statemap[c_i\leftarrow q'_i]_{i \in \interv{1}{n}}$ and
$q_i \arrow{p_i}{} q'_i$ is a transition in $\beh$, for each
$i \in \interv{1}{n}$. For all $i \in \interv{1}{n}$, since
$c_i \in \comps$, there exists a unique node $w_i \in \dom{\tree}$ and
a variable $\tavarcomp{i} \in \set{x_{k_i}^{w_i}, y_{\ell_i}^{w_i}}$,
for some $k_i$, $\ell_i \geq 1$, such that $\store'(\tavarcomp{i}) =
c_i$ and $\compactin{\tavarcomp{i}}{q_i}$ occurs in
$\formulaoftreefree^\epsilon(\tree)$, for some $q_i \in \states$. Then
the tree $\tree'$ is defined as $\dom{\tree'} = \dom{\tree}$
and: \begin{compactitem}
\item $\tree'(w_i) = \tuple{\psi^{w_i},a^{w_i}_0,\ldots,a^{w_i}_h}$, where
      $\psi^{w_i}$ is obtained from $\phi^{w_i}$ by replacing the atom
      $\compactin{\tavarin{k_i}}{q_i}$
      (resp. $\compactin{y_{k_i}}{q_i}$) with
      $\compactin{\tavarin{k_i}}{q_i'}$
      (resp. $\compactin{y_{k_i}}{q_i'}$),
\item $\tree'(u)=\tree(u)$, for all $u \in \dom{\tree} \setminus \set{w_1, \ldots, w_n}$.  
\end{compactitem}
The check of
$\aconfig' \models^{\store'} \formulaoftreefree^\epsilon(\tree')$
follows from the definition of $\tree'$ and
$\statemap'=\statemap[c_i\leftarrow q'_i]_{i \in \interv{1}{n}}$.

\noindent(2) We prove that $(\tree,\tree') \in \lang{\trans_{(p_1,\ldots,p_n)}}$
by building an accepting run $\pi
: \dom{\tree} \finmap \tastates_{(p_1,\ldots,p_n)}$ of
$\trans_{(p_1,\ldots,p_n)}$ over $(\tree,\tree')$. For each node
$u \in \dom{\tree}$, the formula $\pi(u)$ is the separating
conjunction of the following equality atoms: \begin{compactitem}
\item $\tavarin{i} = \tavarin{j}$, where $\formulaoftreefree^u(\subtree{\tree}{u}) \models x_{i}^u = x_{j}^u$,
  for all $i, j \in \interv{1}{a^u_0}$, 
\item $\tavarin{j} = \tavarbegin{i}$, where $\formulaoftreefree^u(\subtree{\tree}{u}) \models x_{j}^u = \tavarcomp{i}$,
  for all $i \in \interv{1}{n}$ and $j \in \interv{1}{a^u_0}$, 
\item $\tavarin{j} = \tavarend{i}$, where $\formulaoftreefree^u(\subtree{\tree}{u}) \models x_{j}^u = \tavarinter{i}$,
  for all $i \in \interv{1}{n}$ and $j \in \interv{1}{a^u_0}$, 
\item $\tavarbegin{i} = \tavarend{i}$, where $\formulaoftreefree^u(\subtree{\tree}{u}) \models \tavarcomp{i} = \tavarinter{i}$,
  for all $i \in \interv{1}{n}$.
\end{compactitem} 
It is easy to check that $\pi$ is indeed a run of
$\trans_{(p_1,\ldots,p_n)}$, by Def. \ref{def:trans}. We are left with
proving that $\pi$ is accepting i.e., that
$\pi(\epsilon) \in \tafinstates_{(p_1,\ldots,p_n)}$.  Let
$\interacn{\tavarinter{1}}{p_1}{\tavarinter{n}}{p_n}$, where
$\tavarinter{i} \in \set{x^{w_0}_{k_i}, y^{w_0}_{\ell_i}}$, for some
$k_i,\ell_i\geq1$, be the interaction atom of
$\formulaoftreefree^\epsilon(\tree)$, such that
$\store'(\tavarinter{i})=c_i$, for all $i\in\interv{1}{n}$.  Since
$\formulaoftreefree^\epsilon(\tree)$ is satisfiable, there is exactly
one such node in $\dom{\tree}$. Because
$\formulaoftree^\epsilon(\tree)$ is tight, the formula
$\formulaoftreefree^\epsilon(\tree)$ is tight and, by
Lemma \ref{lemma:walks}, there exists an equality walk between the
nodes $w_i$ and $w_0$, for all $i \in \interv{1}{n}$. By the
definition of $\pi$, we obtain $\pi(\epsilon) \models \tavarbegin{i}
= \tavarend{i}$ for all $i\in\interv{1}{n}$, thus
$\pi(\epsilon) \in \tafinstates_{(p_1,\ldots,p_n)}$.

\end{proofE}
Note that the condition of $\formulaoftree^\epsilon(\tree)$ having
only tight models is necessary to avoid interactions
$(c_i,p_i)_{i\in\interv{1}{n}}$ that fire by ``accident'' i.e., when
the interaction is created by an atom
$\interacn{\tavarinter{1}}{p_1}{\tavarinter{n}}{p_n}$, with the
components $c_1, \ldots, c_n$ created by component atoms
$\compactin{\tavarcomp{1}}{q_1}, \ldots,
\compactin{\tavarcomp{n}}{q_n}$, such that the equality $\tavarcomp{i}
= \tavarinter{i}$ is not the consequence of
$\formulaoftree^\epsilon(\tree)$, for some $i \in \interv{1}{n}$. The
effect of such interactions is not captured by the transducer
introduced by Def. \ref{def:trans}. Tightness is, moreover, a
necessary condition of Lemma \ref{lemma:walks}, that ensures the
existence of equality walks between the variables occurring in an
interaction atom and those of the atoms creating the components to
which these variables are mapped, in a model of
$\formulaoftree^\epsilon(\tree)$.


\begin{lemmaE}\label{lemma:semtransleft}
  For each tree $\tree' \in \trans_{(p_1,\ldots,p_n)}(L)$,
  configuration $\aconfig' \in \configset$ and store $\store$, such
  that $\aconfig' \models^\store \formulaoftree^\epsilon(\tree')$,
  there exists a configuration $\aconfig =
  (\comps,\interacs,\statemap)$ and a tree $\tree \in L$, such that
  $\aconfig \models^\store \formulaoftree^\epsilon(\tree)$ and
  $\aconfig \step{(c_i,p_i)_{i \in \interv{1}{n}}} \aconfig'$, for
  some $c_1, \ldots, c_n \in \comps$ and $(c_i,p_i)_{i \in
    \interv{1}{n}} \in \interacs$.
\end{lemmaE}
\begin{proofE}
  Let $\aconfig' \isdef (\comps,\interacs,\statemap')$. Because
$\aconfig' \models^\store \formulaoftree^\epsilon(\tree')$, by
Def. \ref{def:charform}, there exists a store $\store'$, such that
$\aconfig' \models^{\store'} \formulaoftreefree^\epsilon(\tree')$. Moreover,
since $\tree' \in \trans_{(p_1,\ldots,p_n)}(L)$, there exists a tree
$\tree \in L$, such that $\dom{\tree} = \dom{\tree'}$,
$(\tree,\tree') \in \lang{\trans_{(p_1,\ldots,p_n)}}$ and let $\pi
: \dom{\tree} \rightarrow \tastates_{(p_1,\ldots,p_n)}$ be an
accepting run of $\trans_{(p_1,\ldots,p_n)}$ over $(\tree,\tree')$. By
Def. \ref{def:trans}, for each $u \in \dom{\tree}$, the state $\pi(u)$
is the separating conjunction of all equalities entailed by
$\formulaoftreefree^u(\subtree{\tree}{u})$. Since $\pi$ is accepting,
we have
$\pi(\epsilon) \models \Asterisk_{i\in\interv{1}{n}} \tavarbegin{i}
= \tavarend{i}$. By Def. \ref{def:trans}, there exists nodes
$u,v \in \dom{\tree}$ and variables $\tavarcomp{i}$ and
$\tavarinter{i}$, such that, for all $i \in \interv{1}{n}$, the
component atom $\compactin{\tavarcomp{i}}{q_i}$
(resp. $\compactin{\tavarcomp{i}}{q'_i}$) occurs in $\tree(u)$
(resp. $\tree'(u)$), for a transition $q_i \arrow{p_i}{} q'_i$ of
$\beh$, and the interaction atom
$\interacn{\tavarinter{1}}{p_1}{\tavarinter{n}}{p_n}$ occurs in both
$\tree(v)$ and in $\tree'(v)$. We define the state map $\statemap
= \statemap'[\store'(\tavarcomp{i}) \leftarrow
q_i]_{i \in \interv{1}{n}}$ and let $\aconfig =
(\comps,\interacs,\statemap)$. We are left with proving that
$\aconfig \models \formulaoftree^\epsilon(\tree)$ and
$\aconfig \step{(\store'(\tavarcomp{i}),p_i)_{i \in \interv{1}{n}}} \aconfig'$,
which are both easy checks.

\end{proofE}


\subsection{The Main Result}
\label{sec:main}

We establish the main result of this section, which is a many-one
reduction of the havoc invariance to the entailment problem. The
result is sharpened by proving that the
reduction \begin{inparaenum}[(i)]
\item preserves the class of the SID (see Def. \ref{def:class} below),
  and
\item is polynomial when several parameters of the SID are bounded by
  constants and simply exponential otherwise.
\end{inparaenum}
In particular, a class-preserving polynomial reduction ensures that
the decidability and complexity upper bounds of the entailment problem
carry over to the havoc invariance problem.

\begin{definition}\label{def:class}
  For two predicate-free formul{\ae} $\phi$ and $\psi$, we write $\phi
  \simeq \psi$ if and only if they become equivalent when dropping the
  state atoms from both. For an arity-preserving equivalence relation
  $\sim~ \subseteq \preds \times \preds$ (i.e.,
  $\arityof{\apred}=\arityof{\bpred}$, for all $\apred\sim\bpred$),
  for any two rules $\arule_1$ and $\arule_2$, we write $\arule_1
  \approx \arule_2$ if and only if $\arule_1 = \apred(x_1, \ldots,
  x_{\arityof{\apred}}) \leftarrow \exists y_1 \ldots \exists y_m ~.~
  \phi * \Asterisk_{\ell\in\interv{1}{h}} \bpred_\ell(\vec{z}_\ell)$,
  $\arule_2=\apred'(x_1, \ldots, x_{\arityof{\apred'}}) \leftarrow
  \exists y'_1 \ldots \exists y'_p ~.~ \psi *
  \Asterisk_{\ell\in\interv{1}{h}} \bpred'_\ell(\vec{u}_\ell)$,
  $\exists y_1 \ldots \exists y_m ~.~ \phi \simeq \exists y'_1 \ldots
  \exists y'_p ~.~ \psi$, $\apred \sim \apred'$ and $\bpred_\ell \sim
  \bpred'_\ell$, for all $\ell \in \interv{1}{h}$. For two SIDs
  $\asid_1$ and $\asid_2$, we write $\asid_1 \preceq \asid_2$ if and
  only if for each rule $\arule_1 \in \asid_1$ there exists a rule
  $\arule_2 \in \asid_2$, such that $\arule_1 \approx \arule_2$. We
  denote by $\asid_1 \approx \asid_2$ the conjunction of $\asid_1
  \preceq \asid_2$ and $\asid_2 \preceq \asid_1$.
\end{definition}
If $\apred_1 \sim \apred_2$ and $\asid_1 \approx \asid_2$ then
$\asid_1$-models of $\apred_1(x_1, \ldots, x_{\arityof{\apred_1}})$
differ from the $\asid_2$-models of $\apred_2(x_1, \ldots,
x_{\arityof{\apred_1}})$ only by a renaming of the states occurring
within state atoms. This is because any derivation of the satisfaction
relation $\aconfig \models_{\asid_1}^\store \apred_1(x_1, \ldots,
x_{\arityof{\apred_1}})$ can be mimicked (modulo the state
atoms that may change) by a derivation of $\aconfig \models_{\asid_2}^\store
\apred_2(x_1, \ldots, x_{\arityof{\apred_2}})$, and viceversa. We are
now in the position of stating the main result of this section:

\begin{theoremE}\label{thm:reduction}
  Assuming that $\apred(x_1, \ldots, x_{\arityof{\apred}})$ is a
  $\asid$-tight formula, each instance $\hinv{\asid}{\apred}$ of the
  havoc invariance problem can be reduced to a set
  $\Set{\overline{\apred}_i(x_1, \ldots, x_{\arityof{\apred}})
    \models_{\asid \cup \overline{\asid}} \apred(x_1, \ldots,
    x_{\arityof{\apred}})}_{i=1}^p$ of entailments, where $\asid
  \approx \overline{\asid}$, for an arity-preserving equivalence
  relation $\sim~ \subseteq \preds \times \preds$, such that
  $\overline{\apred}_i \sim \apred$, for all $i \in
  \interv{1}{p}$. The reduction is polynomial, if
  $\maxarityof{\asid}$, $\maxinterof{\asid}$ and $\maxpredsof{\asid}$
  are bounded by constants and simply exponential, otherwise.
\end{theoremE}
\begin{proofE}
  Let $\hinv{\asid}{\apred}$ be an instance of the havoc invariance
  problem. We denote by $\intertypeof{\asid}$ the set of types $\tau =
  (p_1, \ldots, p_n)$ of an interaction atom
  $\interacn{y_1}{p_1}{y_n}{p_n}$ that occurs in $\asid$. Let
  $\auth_\asid$ be the tree automaton from Def. \ref{def:sid-ta} and let: 
  \[\trans = \Big(\talpha^2_\asid, \bigcup_{\scriptstyle{\tau \in
    \intertypeof{\asid}}} \tastates_{\intertype},
  \bigcup_{\scriptstyle{\tau \in \intertypeof{\asid}}}
  \tafinstates_{\intertype}, \bigcup_{\scriptstyle{\tau \in
      \intertypeof{\asid}}} \tatrans_{\intertype}\Big)\] be the
  automata-theoretic union of the transducers $\trans_\tau$, taken
  over all $\tau \in \intertypeof{\asid}$. Let $\overline{\auth}$ be
  the tree automaton that recognizes the language
  $\trans(\langof{q_\apred}{\auth_\asid})$. The states of
  $\overline{\auth}$ are pairs of the form $(q_\bpred,\varphi)$, where
  $\bpred \in \defnof{\asid}$ and $\varphi \in \tastates_\tau$, for
  some $\tau \in \intertypeof{\asid}$. Moreover, the final states of
  $\overline{\auth}$ are of the form $(q_\apred,\varphi)$, where
  $\varphi \in \tafinstates_\tau$, for some $\tau \in
  \intertypeof{\asid}$. Let $\overline{\asid} \isdef
  \asid_{\overline{\auth}}$ be the SID from Def. \ref{def:ta-sid}
  relative to $\overline{\auth}$ and let $\set{\overline{\apred}_i
    \mid i \in \interv{1}{p}} \isdef \set{\apred_{(q_\apred,\varphi)}
    \mid \varphi \in \tafinstates_\tau,~ \tau \in
    \intertypeof{\asid}}$ be the set of predicates corresponding to
  the final states of $\overline{\auth}$. It is easy to check that
  $\overline{\auth}$ is SID-compatible (Def. \ref{def:sid-compatible})
  and that $\arityof{\overline{\apred}_i} = \arityof{\apred}$, for all
  $i \in \interv{1}{p}$. For a store $\store$, we denote by
  $\store[x_i/y_i]_{i \in \interv{1}{n}}$ the store that maps $x_i$ to
  $\store(y_i)$ and agrees with $\store$ everywhere else. We prove
  that $\hinv{\asid}{\apred}$ has a positive answer if and only if
  $\overline{\apred}_i(x_1, \ldots, x_{\arityof{\apred}})
  \models_{\overline{\asid}\cup\asid} \apred(x_1, \ldots,
  x_{\arityof{\apred}})$ holds, for all $i \in \interv{1}{p}$.

  \vspace*{\baselineskip}\noindent ``$\Rightarrow$'' Let $\aconfig'$
  be a configuration and $\store$ be a store, such that $\aconfig'
  \models^\store_{\overline{\asid}} \overline{\apred}_i(x_1, \ldots,
  x_{\arityof{\apred}})$, for some $i \in \interv{1}{p}$. Let $\store'
  \isdef \store[x^\epsilon_i / x_i]_{i \in
    \interv{1}{\arityof{\apred}}}$ be a store, such that $\aconfig'
  \models_{\overline{\asid}} \overline{\apred}_i(x^\epsilon_1, \ldots,
  x^\epsilon_{\arityof{\apred}})$. By Lemma \ref{lemma:ta-sid}, there
  exista a tree
  $\tree'\in\langof{(q_\apred,\varphi)}{\overline{\auth}}$, for a
  final state $(q_\apred,\varphi)$ of $\mathcal{T}_{(p_1, \ldots,
    p_n)}$, for some interaction type $(p_1, \ldots, p_n) \in
  \intertypeof{\asid}$, such that $\aconfig' \models^{\store'}
  \formulaoftree^\epsilon(\tree')$. By Lemma \ref{lemma:semtransleft},
  there exists a tree $\tree \in \langof{q_\apred}{\auth_\asid}$ and a
  configuration $\aconfig$, such that $\aconfig \models^\store
  \formulaoftree^\epsilon(\tree)$ and $\aconfig \step{(c_i,p_i)_{i \in
      \interv{1}{n}}} \aconfig'$, for some interaction $(c_i,p_i)_{i
    \in \interv{1}{n}}$ from $\aconfig$. By Lemma \ref{lemma:sid-ta},
  we obtain that $\aconfig \models_\asid^{\store'}
  \apred(x^\epsilon_1, \ldots, x^\epsilon_{\arityof{\apred}})$,
  leading to $\aconfig \models_\asid^{\store} \apred(x_1, \ldots,
  x_{\arityof{\apred}})$. By the hypothesis that
  $\hinv{\asid}{\apred}$ has a positive answer, we obtain that
  $\aconfig' \models^\store_\asid
  \apred(x_1,\ldots,x_{\arityof{\apred}})$. Since the choices of
  $\aconfig'$ and $i$ were arbitrary, we obtain
  $\overline{\apred}_i(x_1, \ldots, x_{\arityof{\apred}})
  \models_{\overline{\asid} \cup \asid} \apred(x_1, \ldots,
  x_{\arityof{\apred}})$, for all $i \in \interv{1}{p}$.

  \vspace*{\baselineskip}\noindent ``$\Leftarrow$'' Let $\aconfig$,
  $\aconfig'$ be configurations and $\store$ be a store, such that
  $\aconfig \models^\store \apred(x_1, \ldots, x_{\arityof{\apred}})$
  and $\aconfig \step{(c_i,p_i)_{i \in \interv{1}{n}}} \aconfig'$, for
  some interaction $(c_i,p_i)_{i \in \interv{1}{n}}$ from $\aconfig$.
  Let $\store' \isdef \store[x^\epsilon_i / x_i]_{i \in
    \interv{1}{\arityof{\apred}}}$ be a store such that $\aconfig
  \models_\asid^{\store'} \apred(x^\epsilon_1, \ldots,
  x^\epsilon_{\arityof{\apred}})$. By Lemma \ref{lemma:sid-ta}, there
  exists a tree $\tree \in \langof{q_\apred}{\auth_\asid}$, such that
  $\aconfig \models^{\store'} \formulaoftree^\epsilon(\tree)$. By the
  assumption in the statement, since $\apred(x_1, \ldots,
  x_{\arityof{\apred}})$ and hence $\apred(x^\epsilon_1, \ldots,
  x^\epsilon_{\arityof{\apred}})$ is $\asid$-tight, and every model of
  $\formulaoftree^\epsilon(\tree)$ is an $\asid$-model of
  $\apred(x^\epsilon_1, \ldots, x^\epsilon_{\arityof{\apred}})$, we
  obtain that $\formulaoftree^\epsilon(\tree)$ is tight. Since
  $\aconfig \step{(c_i,p_i)_{i \in \interv{1}{n}}} \aconfig'$, there
  exists a tree $\tree' \in \trans_{(p_1, \ldots,
    p_n)}(\langof{q_\apred}{\auth_\asid}) =
  \langof{(q_\apred,\varphi)}{\overline{\auth}}$, such that $\aconfig'
  \models^{\store'} \formulaoftree^\epsilon(\tree')$, where $\varphi$
  is a final state of $\trans_{(p_1, \ldots, p_n)}$, by Lemma
  \ref{lemma:semtransright}. By the definition of $\overline{\asid}$
  and of the predicates $\overline{\apred}_1, \ldots,
  \overline{\apred}_p$, by Lemma \ref{lemma:sid-ta}, we obtain
  $\aconfig' \models^{\store'}_{\overline{\asid}}
  \overline{\apred}_i(x^\epsilon_1, \ldots,
  x^\epsilon_{\arityof{\apred}})$, for some $i \in \interv{1}{p}$. We
  obtain $\aconfig' \models^{\store}_{\overline{\asid}}
  \overline{\apred}_i(x_1, \ldots, x_{\arityof{\apred}})$ and the
  hypothesis $\overline{\apred}_i(x_1, \ldots, x_{\arityof{\apred}})
  \models_{\overline{\asid} \cup \asid} \apred(x_1, \ldots,
  x_{\arityof{\apred}})$ yields $\aconfig' \models^\store_\asid
  \apred(x_1, \ldots, x_{\arityof{\apred}})$. Since the choices of
  $\aconfig$ and $\aconfig'$ were arbitrary, we obtain that
  $\hinv{\asid}{\apred}$ has a positive answer.

  \vspace*{\baselineskip} To prove $\asid \approx \overline{\asid}$,
  consider the relation $\bpred \sim \apred_{(q_\bpred,\varphi)}$, for
  all $\bpred \in \preds$ and all states $\varphi \in \bigcup_{\tau
    \in \intertypeof{\asid}} \tastates_{\intertype}$. Because
  $\overline{\auth}$ is SID-compatible, we have $\arityof{\bpred} =
  \arityof{\apred_{(q_\bpred,\varphi)}}$, for all $\bpred \in \preds$,
  hence $\sim$ is arity-preserving. By the definition of
  $\overline{\asid}$, for each rule $\bpred_0(x_1, \ldots,
  x_{\arityof{\bpred_0}}) \leftarrow \exists y_1 \ldots \exists y_m ~.~
  \phi * \Asterisk_{\ell=1}^h \bpred_\ell(z^\ell_1, \ldots,
  z^\ell_{\arityof{\bpred_\ell}})$ from $\asid$ there exists a rule:
  \begin{align}
    \apred_{(q_{\bpred_0},\varphi_0)} \leftarrow \exists y'_1 \ldots \exists y'_p ~.~
    \psi * \Asterisk_{\ell=1}^h \apred_{(q_{\bpred_\ell,\varphi_\ell})}(u^\ell_1, \ldots,
    u^\ell_{\arityof{\bpred_\ell}}) \label{eq:rule}
  \end{align} in $\overline{\asid}$, such that
  $\exists y_1 \ldots \exists y_m ~.~ \phi \simeq \exists y'_1 \ldots
  \exists y'_p ~.~ \psi$. The last equivalence between predicate-free
  formul{\ae} is because $\exists y'_1 \ldots \exists y'_p ~.~ \psi$ is
  obtained from $\exists y_1 \ldots \exists y_m ~.~ \phi$ by changing
  state atoms and introducing equality atoms of which one of the
  variable is existentially quantified (see Def. \ref{def:sid-ta},
  Def. \ref{def:trans} and Def. \ref{def:ta-sid}). Viceversa, each rule
  of the form (\ref{eq:rule}) in $\overline{\asid}$ stems from a rule
  in $\asid$, where the same equivalences hold. We conclude that
  $\asid \approx \overline{\asid}$, by Def. \ref{def:class}.

  \vspace*{\baselineskip} Finally, we compute an upper bound on the
  time necessary to build $\overline{\asid}$ from $\asid$. Note that
  the number of interaction atoms, and hence the number of interaction
  types $\tau \in \intertypeof{\asid}$, that occur in $\asid$ is
  bounded by $\sizeof{\asid}$. The number of states in each transducer
  $\trans_\tau$, for some $\tau \in \intertypeof{\asid}$, is bounded
  by the number of partitions of
  $\interv{1}{\maxarityof{\asid}+\maxinterof{\asid}}$, that is
  $2^{\bigO((\maxarityof{\asid}+\maxinterof{\asid}) \cdot
    \log(\maxarityof{\asid}+\maxinterof{\asid})}$. This is because
  each state $\eqformulae{\ttvars{\tau}}$ corresponds (modulo logical
  equivalence) to a partition of the set $\set{\tavarin{1}, \ldots,
    \tavarin{\arityof{\apred}}} \cup \set{\tavarbegin{i}, \tavarend{i}
    \mid i \in \interv{1}{\maxinterof{\asid}}}$ of the parameters of
  some predicate $\apred \in \defnof{\asid}$ to which the variables
  $\tavarbegin{i}, \tavarend{i}$, for $i = 1, \ldots, n$ are added and
  the number of partitions of this set is aymptotically bounded by
  $(\arityof{\apred}+n)^{(\arityof{\apred}+n)}$.

  The size of the transducer alphabet is the number of pairs
  $(\alpha,\beta)$, such that $\alpha = \tuple{\phi, a_0, \ldots,
    a_h}$, $\beta = \tuple{\psi, a_0, \ldots, a_h}$, $\phi$ stems from
  a rule in $\asid$ and $\psi$ differs from $\phi$ by a renaming of
  states in at most $n$ state atoms, for $n \leq
  \maxinterof{\asid}$. Let $M$ be the maximum number of variables that
  occurs free or bound in a rule in $\asid$ and $B$ be the maximum
  number of outoing transitions $q \arrow{p}{} q'$ in the behavior
  $\beh = (\ports,\states,\arrow{}{})$. Clearly $M \leq
  \sizeof{\asid}$, whereas $B$ is a constant, because $\beh$ is
  considered to be fixed i.e., not part of the input of the havoc
  invariance problem. Then, for each \qpf\ formula that occurs in
  $\asid$, the number of alphabet symbols of the form
  $(\tuple{\phi,a_0,\ldots,a_h},\tuple{\psi,a_0,\ldots,a_h})$ is at
  most:
  \[\begin{array}{rcl}
  \sum_{i=1}^{\maxinterof{\asid}}\binom{M}{i} \cdot B^i & \leq & B^{\maxinterof{\asid}} \cdot \sum_{i=0}^{\maxinterof{\asid}}\binom{M}{i} \\
  & \leq & B^{\maxinterof{\asid}} \cdot \sum_{i=0}^{\maxinterof{\asid}} \frac{M^i}{i!} = B^{\maxinterof{\asid}} \cdot \sum_{i=0}^{\maxinterof{\asid}} \frac{\maxinterof{\asid}^i}{i!} \cdot \left(\frac{M}{\maxinterof{\asid}}\right)^i \\
  & \leq & B^{\maxinterof{\asid}} \cdot  \left(\frac{M}{\maxinterof{\asid}}\right)^{\maxinterof{\asid}} \cdot \sum_{i=0}^{\maxinterof{\asid}} \frac{\maxinterof{\asid}^i}{i!} \\
  & \leq & \left(\frac{B \cdot M \cdot e}{\maxinterof{\asid}}\right)^{\maxinterof{\asid}} \leq (B \cdot \sizeof{\asid})^{\maxinterof{\asid}} 
  \end{array}\]
  Given alphabet symbols $\alpha = \tuple{\phi,a_0,\ldots,a_h}$,
  $\beta = \tuple{\psi,a_0,\ldots,a_h}$ and states $\varphi_0, \ldots,
  \varphi_h$, the time required to decide on the existence of a
  transition $(\alpha,\beta)(\varphi_1, \ldots, \varphi_h) \rightarrow
  \varphi_0$ is $\bigO((\size{\phi} + \sum_{i=0}^h
  \size{\varphi_i})^3) = \bigO(\size{\asid}^3 \cdot
  (\maxarityof{\asid} + \maxinterof{\asid})^3) =
  \bigO(\size{\asid}^3)$, because eliminating existential quantifiers
  from and checking equivalence between separating conjunctions of
  equality atoms is cubic in the number of variables, using a standard
  Floyd-Warshall closure algorithm. Summing up, the time needed to
  compute the transducer $\trans$ as the union of
  $\trans_{(p_1,\ldots,p_n)}$, for all $(p_1,\ldots,p_n) \in
  \intertypeof{\asid}$ is bounded by:
  \[\begin{array}{l}
  \sizeof{\asid}^{\maxinterof{\asid}+3} \cdot
  2^{\bigO(\maxpredsof{\asid} \cdot (\maxarityof{\asid}+\maxinterof{\asid}) \cdot \log(\maxarityof{\asid}+\maxinterof{\asid})}
  \end{array}\]
  This is because there are at most $(B \cdot
  \sizeof{\asid})^{\maxinterof{\asid}} \cdot \cardof{\tastates_{(p_1,
      \ldots, p_n)}}^{\maxpredsof{\asid}+1}$ transitions of the form
  $(\alpha,\beta)(q_1,\ldots,q_h) \arrow{}{} q_0$ in
  $\trans_{(p_1,\ldots,p_n)}$ and their enumeration requires time
  simply exponential in $\maxpredsof{\asid}$. Since the translation
  between a SID and a tree automaton takes linear time, the above is
  an upper bound for the reduction of the $\hinv{\asid}{\apred}$
  instance of the havoc invariance problem to the set
  $\Set{\overline{\apred}_i(x_1, \ldots, x_{\arityof{\apred}})
    \models_{\asid \cup \overline{\asid}} \apred(x_1, \ldots,
    x_{\arityof{\apred}})}_{i=1}^p$ of instances of the entailment
  problem. The reduction is thus polynomial, if $\maxarityof{\asid}$,
  $\maxinterof{\asid}$ and $\maxpredsof{\asid}$ are constant and
  simply exponential, otherwise.
\end{proofE}

\section{Decidability and Complexity}
\label{sec:complexity}

We prove the undecidability of the havoc invariance problem
(Def. \ref{def:havoc-invariance}) using a reduction from the
universality of context-free languages, a textbook undecidable
problem \cite{BarHillel61}.

\ifLongVersion
The idea of the reduction is to encode a word as a chain of components
connected with binary interactions between left and right neighbours
(see \S\ref{sec:motivating-example} for the inductive definition of a
chain of components). The symbols in the word are encoded by the type
(i.e. tuple of ports) of the interactions. The behavior consists of
transitions $q_0 \arrow{p}{} q_1$ and $q_1 \arrow{p}{} q_1$, for each
port $p \in \ports$. A chain corresponding to a word from $\Sigma^*$
has exactly one component in state $q_0$, hence any sequence of
interactions in the chain may change the state of at most one
component. The productions of $G$ are encoded by rules of the SID
$\asid$ and the words produced by $G$ are chains with every component
in state $q_1$. Finally, a predicate symbol $\apred$ describes the
union of the sets of configurations corresponding to words in
$\Sigma^*$ and $\lang{G}$, respectively. Then
$\Sigma^* \subseteq \lang{G}$ if and only if each configuration
encoding a word from $\Sigma^*$, following the execution of zero or
more interactions that change the state of one of its components from
$q_0$ to $q_1$, encodes a word from $\lang{G}$ i.e., if and only if
$\hinv{\asid}{\apred}$ has a positive answer.
\fi

\begin{theoremE}\label{thm:undecidability}
  The $\hinv{\asid}{\apred}$ problem is undecidable. 
\end{theoremE}
\begin{proofE}
    A \emph{context-free grammar} $G = (\Sigma,N,T,S,\Delta)$ consists
        of a finite set $N$ of \emph{nonterminals}, a finite set $T$
        of words over a finite alphabet $\Sigma$,
        called \emph{terminals}, a start symbol $S \in N$ and a finite
        set $\Delta$ of \emph{productions} of the form $A \rightarrow
        w$, where $A \in N$ and $w \in (N \cup T)^*$. Given finite
        strings $u, v \in (N \cup T)^*$, the relation $u \rhd v$
        replaces a nonterminal $A$ of $u$ by the right-hand side $w$
        of a production $A \rightarrow w$ and $\rhd^*$ denotes the
        reflexive and transitive closure of
        $\rhd$. The \emph{language} of $G$ is the set $\lang{G}$ of
        finite strings $w \in T^*$, such that $S \rhd^*
        w$. The \emph{universality problem} asks, given a grammar $G$,
        whether $\Sigma^* \subseteq \lang{G}$? Let $G =
        (\Sigma,N,T,S,\Delta)$ be a context-free grammar and assume
        w.l.o.g. the following: \begin{compactitem}
  \item $\Sigma = \set{0,1}$, because every symbol can be encoded as a
    binary string,
  \item $G$ does not produce the empty word or the single letter words
    $0$ and $1$; computing a grammar $G'$ such that $\lang{G'} =
    \lang{G} \setminus \set{\epsilon,0,1}$ is possible and we can
    reduce from the modified universality problem $\Sigma^{\geq2}
    \isdef \set{w \in \Sigma^* \mid \lenof{w} \geq 2} \subseteq
    \lang{G'}$ instead of the original problem $\Sigma^* \subseteq
    \lang{G}$.
  \item $G$ is in Greibach normal form i.e., contains only production
    rules of the form $Y_0 \rightarrow a Y_1 \ldots Y_k$, where $Y_0,
    \ldots, Y_k \in N$, for some $k \geq 0$ and $a \in \Sigma$.
  \end{compactitem}  
  We define the behavior $\beh=(\set{p},\set{q_0,q_1},\arrow{}{})$,
  where $q_0 \arrow{p}{} q_1$ and $q_1 \arrow{p}{} q_1$. We encode the
  language $\Sigma^{\geq2}$ by the following set of rules, that define
  the binary predicate symbols $\mathsf{X}_0(x,y)$ and
  $\mathsf{X}_1(x,y)$ below:
  \[\begin{array}{rcl}
  \mathsf{X}_0(x,y) & \leftarrow & \exists z ~.~ \compactin{x}{q_0}
  * \interactwo{x}{p}{z}{p} * \mathsf{X}_1(z,y) \\
  \mathsf{X}_0(x,y) & \leftarrow & \exists z ~.~ \compactin{x}{q_1} * \interactwo{x}{p}{z}{p} * \mathsf{X}_0(z,y) \\
  \mathsf{X}_1(x,y) & \leftarrow & \compactin{x}{q_1} * x=y \\
  \mathsf{X}_1(x,y) & \leftarrow & \exists z ~.~ \compactin{x}{q_1} * \interactwo{x}{p}{z}{p} * \mathsf{X}_1(z,y)
  \end{array}\]
  Note that $\mathsf{X}_0(x,y)$ defines those configurations encoding
  words of length at least two, with exactly one component in state
  $q_0$ and the rest of the components in state $q_1$. To encode the
  language $\lang{G}$, we use a binary predicate symbol
  $\mathsf{Y}_i(x,y)$, where $i \in \interv{1}{n}$, for each
  nonterminal from the set $N = \set{Y_1, \ldots, Y_n}$ and encode
  each production rule of $G$ of the form $Y_{i_0} \rightarrow a
  Y_{i_1} \ldots Y_{i_k}$, for some $k \geq 0$, by a
  rule:
  \[\begin{array}{rcl}
  \mathsf{Y}_{i_0}(x,y) & \leftarrow & \exists z_1 \ldots \exists z_k ~.~ \compactin{x}{q_1} * \interac{x}{p}{z_1}{p}
  * \Asterisk_{\ell=1}^{k-1} \mathsf{Y}_{i_\ell}(z_\ell,z_{\ell+1}) * \mathsf{Y}_{i_k}(z_k,y)
  \end{array}\]
  Note that $\mathsf{Y}_i(x,y)$ encodes the words produced by $G$ starting with
  nonterminal $Y_i$, by a chain configurations, in which all
  components are in a state $q_1$. In particular, the parameter $y$ is
  not bound to an present component, allowing to compose
  $\mathsf{Y}_i$ with other predicate atoms $\mathsf{Y}_j$.  Finally,
  we define the predicate $\apred$ by the following
  rules:
  \[\begin{array}{rcl}
  \apred(x,y) & \leftarrow & \mathsf{X}_0(x,y) \\
  \apred(x,y) & \leftarrow & \mathsf{Y}_0(x,y) * \compactin{y}{q_1}
  \end{array}\]
  assuming w.l.o.g. that $Y_0$ is the starting nonterminal of $G$.
  Let $\asid$ be the SID consisting of the rules above. We prove that
  $\Sigma^{\geq2} \subseteq \lang{G}$ if and only if for any
  configurations $\aconfig, \aconfig'$ and each store $\store$, such
  that $\aconfig \models^\store_\asid \apred(x,y)$ and
  $\aconfig \Havoc \aconfig'$, we have
  $\aconfig' \models^\store_\asid \apred(x,y)$. ``$\Rightarrow$'' Let
  $\aconfig,\aconfig'$ and $\store$ be such that
  $\aconfig \models^\store_\asid \apred(x,y)$ and
  $\aconfig \Havoc \aconfig'$. If $\aconfig=\aconfig'$ we are done, so
  we assume $\aconfig=(\comps,\interacs,\statemap)$ and
  $\aconfig'=(\comps,\interacs,\statemap')$, where
  $\statemap \neq \statemap'$. By the definition of $\beh$, the only
  possibility is that $\statemap' = \statemap[c\leftarrow q_1]$, for
  some component $c\in\comps$, such that $\statemap(c)=q_0$. But then
  $\aconfig \models^\store_\asid \mathsf{X}_0(x,y)$. Since $\aconfig$
  encodes a word $w\in\Sigma^{\geq2}$, we have $w \in \lang{G}$, by
  the hypothesis, hence
  $\aconfig' \models_\asid^\store \mathsf{Y}_0(x,y)
  * \compactin{y}{q_1}$ and
  $\aconfig' \models_\asid^\store \apred(x,y)$ follows, by the
  definition of $\asid$. ``$\Leftarrow$'' Let $w \in \Sigma^{\geq2}$
  be a word and let $\aconfig = (\comps,\interacs,\statemap)$ be a
  configuration such that $\statemap(c)=q_0$, for some $c \in \comps$
  and $\statemap(c')=q_1$, for all $c'\in\comps\setminus\set{c}$. By
  the definition of $\asid$, we have
  $\aconfig \models_\asid^\store \mathsf{X}_0(x,y)$ for a store
  $\store$, hence $\aconfig \models_\asid^\store \apred(x,y)$. Let
  $\aconfig'=(\comps,\interacs,\statemap[c\leftarrow q_1])$. By the
  hypothesis, we have $\aconfig' \models_\asid^\store \apred(x,y)$,
  hence $\aconfig' \models_\asid^\store \mathsf{Y}_0(x,y)
  * \compactin{y}{q_1}$ is the only possibility. Then we obtain
  $w \in \lang{G}$, again by the definition of $\asid$.
\end{proofE}

The undecidability proof for the havoc invariance problem uses an
argument similar to the one used to prove undecidability of the
entailment problem \cite[Theorem 4]{BozgaBueriIosif22Arxiv}. We
leverage further from this similarity and carve a fragment of
\cl\ with a decidable havoc invariance problem, based on the reduction
from Theorem \ref{thm:reduction}. For self-containment reasons, we
recall the definition of a \cl\ fragment for which the entailment
problem is decidable (see \cite[\S6]{BozgaBueriIosif22Arxiv} for more
details and proofs). This definition relies on three, easily
checkable, syntactic restrictions on the rules of the SID and a
decidable semantic restriction on the models of a predicate atom
defined by the SID. The syntactic restrictions use the notion of
profile:

\begin{definition}\label{def:profile}
  The \emph{profile} of a SID $\asid$ is the pointwise greatest
  function $\profile{\asid} : \preds \rightarrow \pow{\nat}$, mapping each
  predicate $\apred$ into a subset of $\interv{1}{\arityof{\apred}}$,
  such that, for each rule $\apred(x_1, \ldots, x_{\arityof{\apred}})
  \leftarrow \phi$ from $\asid$, each atom $\bpred(y_1, \ldots,
  y_{\arityof{\bpred}})$ from $\phi$ and each $i \in
  \profile{\asid}(\bpred)$, there exists $j \in
  \profile{\asid}(\apred)$, such that $x_j$ and $y_i$ are the same
  variable.
\end{definition}
The profile identifies the parameters of a predicate that are always
replaced by a variable $x_1, \ldots, x_{\arityof{\apred}}$ in each
unfolding of $\apred(x_1,\ldots,x_{\arityof{\apred}})$, according to
the rules in $\asid$; it is computed by a greatest fixpoint iteration,
in polynomial time.

\begin{definition}\label{def:pcr}
  A rule $\apred(x_1, \ldots, x_{\arityof{\apred}}) \leftarrow \exists
  y_1 \ldots \exists y_m ~.~ \phi * \Asterisk_{\ell=1}^h
  \bpred_\ell(z^\ell_1, \ldots, z^\ell_{\arityof{\bpred_\ell}})$,
  where $\phi$ is a \qpf\ formula, is said to
  be: \begin{compactenum}
  \item\label{it:progressing} \emph{progressing} (\textsf{P}) if and only if $\phi
    = \compact{x_1} * \psi$, where $\psi$ consists of interaction
    atoms involving $x_1$ and (dis-)equalities, such that
    $\bigcup_{\ell=1}^h \set{z^\ell_1, \ldots,
      z^\ell_{\arityof{\bpred_\ell}}} = \set{x_2, \ldots,
      x_{\arityof{\apred}}} \cup \set{y_1, \ldots, y_m}$,
  \item\label{it:connected} \emph{connected} (\textsf{C}) if and only if, for each
    $\ell\in\interv{1}{h}$ there exists an interaction atom in $\psi$
    that contains both $z^\ell_1$ and a variable from $\set{x_1} \cup
    \set{x_i \mid i \in \profile{\asid}(\apred)}$,
  \item\label{it:restricted} \emph{equationally-restricted
  (e-restricted or \textsf{R})} if and only if, for every disequality
    $x \neq y$ from $\phi$, we have $\set{x,y} \cap \set{x_i \mid i
    \in \profile{\asid}(\apred)} \neq\emptyset$.
  \end{compactenum}
  A SID $\asid$ is \emph{progressing} (\textsf{P}), \emph{connected}
  (\textsf{C}) and \emph{e-restricted} (\textsf{R}) if and only if
  each rule in $\asid$ is \emph{progressing}, \emph{connected} and
  \emph{e-restricted}, respectively.
\end{definition}

\begin{example}\label{ex:pcr}
For example, the rules for the $\chain{h}{t}(x_1,x_2)$ predicates from
the SID in \S\ref{sec:motivating-example} are \textsf{PCR}, but not the rules
for $\ring{h}{t}()$ predicates, that are neither progressing nor
connected. The latter can be replaced with the following \textsf{PCR} rules:
\vspace*{-.5\baselineskip}
\begin{align*}
  \pcring{h}{t}(x) \leftarrow \exists y \exists z ~.~ \compactin{x}{q} *
  \interactwo{x}{out}{z}{in} * \chain{h'}{t'}(z,y) * \interactwo{y}{out}{x}{in}
  \text{, for all $h, t \in \nat$}
\end{align*}
Similarly, rule (\ref{rule:node}) for the $\treenode$ predicate is
\textsf{PCR}, but not rules (\ref{rule:root}) and (\ref{rule:leaf}),
from Example \ref{ex:tll}. In order to obtain a SID that is
\textsf{PCR}, these rules can be replaced with, respectively:
\vspace*{-.5\baselineskip}
\begin{align*}
  \treeroot(n) \leftarrow & \exists n_1 \exists \ell_1 \exists r_1 \exists n_2 \exists \ell_2 \exists r_2 ~.~
  \compact{n} * \tuple{n.\req,n_1.\reply,n_2.\reply} * \tuple{r_1.\inp, \ell_2.\outp} ~* \nonumber \\
  & \hspace*{4cm} \treenode(n_1,\ell_1,r_1) * \treenode(n_2,\ell_2,r_2) \\
  \treenode(n,\ell,r) \leftarrow & \compact{n} * \tuple{n.\req,\ell.\reply,r.\reply} * \tuple{\ell.\inp,r.\outp} * \treeleaf(\ell) * \treeleaf(r)
  \hspace*{5mm} \treeleaf(n) \leftarrow \compact{n} \text{\hspace*{5mm}$\blacksquare$}
\end{align*}
\end{example}

A first property is that \textsf{PCR} SIDs define only tight
configurations (Def. \ref{def:tightness}), a prerequisite for the
reduction from Theorem \ref{thm:reduction}:

\begin{lemmaE}\label{lemma:pcr-tight}
  Let $\asid$ be a \textsf{PCR} SID and let $\apred \in
  \defnof{\asid}$ be a predicate. Then, for any $\asid$-model
  $(\aconfig, \store)$ of $\apred(x_1, \ldots, x_{\arityof{\apred}})$,
  the configuration $\aconfig$ is tight.
\end{lemmaE}
\begin{proofE}
  Let $\config$ be a configuration and $\store$ be a store, such that
  $\config \models_\asid^\store \apred(x_1, \ldots,
  x_{\arityof{\apred}})$. Let $(c_i, p_i)_{i \in \interv{1}{n}} \in
  \interacs$ be an interaction for which we shall prove that $c_1,
  \ldots, c_n \in \comps$. The proof goes by induction on the
  definition of the satisfaction relation. Assume that \(\config
  \models_\asid^{\store'} \compact{x_1} * \psi * \Asterisk_{\ell=1}^h
  \bpred_\ell(z^\ell_1, \ldots, z^\ell_{\bpred_\ell})\), for a store
  $\store'$ that agrees with $\store$ over $x_1, \ldots,
  x_{\arityof{\apred}}$, where $\psi$ is a separating conjunction of
  interaction atoms. Then there exist configurations $\aconfig_0,
  \ldots, \aconfig_h$, such that $\config = \aconfig_0 \comp \ldots
  \comp \aconfig_h$, $\aconfig_0 \models^{\store'} \compact{x_1} *
  \psi$ and $\aconfig_\ell \models_\asid^{\store'}
  \bpred_\ell(z^\ell_1, \ldots, z^\ell_{\arityof{\bpred_\ell}})$. We
  distinguish two cases: \begin{compactitem}
  \item There exists an interaction atom
    $\interacn{y_1}{p_1}{y_n}{p_n}$ in $\psi$ such that
    $\store'(y_i)=c_i$, for all $i\in\interv{1}{n}$. Since $\asid$ is
    progressing, by Def. \ref{def:pcr}, we have $y_1, \ldots, y_n \in
    \bigcup_{\ell=1}^h \set{z^\ell_1, \ldots,
      z^\ell_{\arityof{\bpred_\ell}}}$. By
    \cite[12]{BozgaBueriIosif22Arxiv}, we obtain $\store'(y_j) \in
    \comps_\ell$, for each $y_j \in \set{z^\ell_1, \ldots,
      z^\ell_{\arityof{\bpred_\ell}}}$, such that $\aconfig_\ell =
    (\comps_\ell, \interacs_\ell, \statemap)$. Consequently, we have
    $\set{c_1, \ldots, c_n} \subseteq \bigcup_{\ell=1}^h \comps_\ell
    \subseteq \comps$, because $\config = \aconfig_0 \comp \ldots
    \comp \aconfig_h$.
  \item Else, the induction hypothesis applies to $\aconfig_\ell
    \models_\asid^{\store'} \bpred_\ell(z^\ell_1, \ldots,
    z^\ell_{\arityof{\bpred_\ell}})$, for some $\ell \in \interv{1}{h}$.
  \end{compactitem}
\end{proofE}

The last restriction for the decidability of entailments relates to
the degree of the models of a predicate atom. The degree of a
configuration is defined in analogy with the degree of a graph as the
maximum number of interactions involving a component:

\begin{definition}\label{def:degree}
  The \emph{degree} of a configuration $\aconfig = (\comps, \interacs,
  \statemap)$ is defined as $\degreeof{\aconfig} \isdef
  \max_{c\in\universe} \degreenode{c}{\aconfig}$, where
  $\degreenode{c}{\aconfig} \isdef \cardof{\{(c_1, p_1, \ldots, c_n,
    p_n) \in \interacs \mid c = c_i,~ i \in \interv{1}{n}\}}$.
\end{definition}
For instance, the configuration of the system from Fig. \ref{fig:ring}
(a) has degree two. The \emph{degree boundedness} problem $\bnd{\asid}{\apred}$ asks, given
a predicate $\apred$ and a SID $\asid$, if the set
$\{\degreeof{\aconfig} \mid \aconfig \models_\asid \exists x_1 \ldots
\exists x_{\arityof{\apred}} ~.~ \apred(x_1, \ldots,
x_{\arityof{\apred}})\}$ is finite. 
\ifLongVersion
\begin{theorem}[\cite{BozgaBueriIosif22Arxiv}]\label{thm:boundedness}
  The $\bnd{\asid}{\apred}$ problem for is in \conp, if
  $\maxarityof{\asid}$ is bounded by a constant, in \exptime, if
  $\maxinterof{\asid}$ is bounded by constant, and in \twoexptime, if
  neither $\maxarityof{\asid}$ nor $\maxinterof{\asid}$ are bounded by
  constants.
\end{theorem}
\else
This problem is decidable \cite[Theorem 3]{BozgaBueriIosif22Arxiv}.
\fi
The entailment problem $\apred(x_1,\ldots,x_{\arityof{\apred}})
\models_\asid \exists x_{\arityof{\apred}+1} \ldots \exists
x_{\arityof{\bpred}} ~.~ \bpred(x_1, \ldots, x_{\arityof{\bpred}})$ is
known to be decidable for \textsf{PCR} SIDs $\asid$, provided,
moreover, that $\bnd{\asid}{\apred}$ holds:

\begin{theorem}[\cite{BozgaBueriIosif22Arxiv}]\label{thm:entailment}
  The entailment problem $\apred(x_1,\ldots,x_{\arityof{\apred}})
  \models_\asid \exists x_{\arityof{\apred}+1} \ldots \exists
  x_{\arityof{\bpred}} ~.~ \bpred(x_1, \ldots, x_{\arityof{\bpred}})$,
  where $\asid$ is \textsf{PCR} and $\bnd{\asid}{\apred}$ has a
  positive answer, is in \twoexptime, if $\maxarityof{\asid}$ and
  $\maxinterof{\asid}$ are bounded by constants and in \fourexptime,
  otherwise.
\end{theorem}

Back to the havoc invariance problem, we give first a lower bound
using a reduction from the entailment problem $\apred(x_1, \ldots,
x_{\arityof{\apred}}) \models_\asid \exists x_{\arityof{\apred}+1}
\ldots \exists x_{\arityof{\bpred}} ~.~ \bpred(x_1, \ldots,
x_{\arityof{\bpred}})$, where $\asid$ is a \textsf{PCR} SID and
$\hinv{\asid}{\apred}$ has a positive answer. To the best of our
efforts, we could not prove that the entailment problem is
\twoexptime-hard under the further assumption that
$\maxarityof{\asid}$ is bounded by a constant, which leaves the
question of a matching lower bound for the havoc invariance problem
open, in this case.

\ifLongVersion Since we can consider, moreover, without loss of
generality, that the rules in $\asid$ contain no state predicates (see
the proof of \cite[Thm. 6]{BozgaBueriIosif22Arxiv} for a reduction
from a fragment of Separation Logic with \twoexptime-complete
entailment problem), we define the SIDs $\asid_0$ and $\asid_1$, such
that \begin{inparaenum}[(1)]
\item the $\asid_0$-models of $\apred(x_1, \ldots,
  x_{\arityof{\apred}})$ have exactly one component in state $q_0$ and
  the rest in state $q_1$,
\item the $\asid_1$-models of $\bpred(x_1, \ldots,
x_{\arityof{\bpred}})$ have each component
in state $q_1$, and
\item each interaction changes both $q_0$ and $q_1$ into $q_1$.
\end{inparaenum}
Further, we introduce a new predicate $\overline{\apred}(x_1, \ldots,
x_{\arityof{\apred}})$ whose models are the $\asid_0$-models of
$\apred(x_1, \ldots, x_{\arityof{\apred}})$ and the $\asid_1$-models
of $\bpred(x_1, \ldots, x_{\arityof{\bpred}})$ and prove that
$\overline{\apred}(x_1, \ldots, x_{\arityof{\apred}})$ is havoc
invariant if and only if the initial entailment $\apred(x_1, \ldots,
x_{\arityof{\apred}}) \models_\asid \exists x_{\arityof{\apred}+1}
\ldots \exists x_{\arityof{\bpred}} ~.~ \bpred(x_1, \ldots,
x_{\arityof{\bpred}})$ holds.
\fi

\begin{lemmaE}\label{lemma:hardness}
  The $\hinv{\asid}{\apred}$ problem for \textsf{PCR} SIDs $\asid$,
  such that $\bnd{\asid}{\apred}$ has a positive answer, is
  \twoexptime-hard.
\end{lemmaE}
\begin{proofE}
  By reduction from the \twoexptime-complete entailment problem
  $\apred(x_1, \ldots, x_{\arityof{\apred}}) \models_\asid \exists
  x_{\arityof{\apred}+1} \ldots \exists x_{\arityof{\bpred}} ~.~
  \bpred(x_1, \ldots, x_{\arityof{\bpred}})$, where $\asid$ is
  \textsf{PCR} and the set $\{\degreeof{\aconfig} \mid \aconfig
  \models_\asid \exists x_1 \ldots \exists x_{\arityof{\apred}} ~.~
  \apred(x_1, \ldots, x_{\arityof{\apred}})\}$ is finite
  \cite[Thm. 4]{BozgaBueriIosif22Arxiv}. Let us fix the above instance
  of the entailment problem and assume w.l.o.g. that there are no
  occurrences of state atoms in $\asid$ --- the proof of the lower
  bound from \cite[Thm. 4]{BozgaBueriIosif22Arxiv} is actually
  independent on the occurrences of state atoms in the rules of the
  SID. Let $\beh = (\ports, \set{q_0,q_1}, \set{q_0 \arrow{p}{} q_1,
    q_1 \arrow{p}{} q_1 \mid p \in \ports})$ be a behavior, where
  $\ports$ is a finite set of ports that subsumes the ports occurring
  in an interaction atom from $\asid$. Moreover, assume w.l.o.g
  that: \begin{compactenum}
  \item\label{ass1} $\apred$ does not occur on the right-hand
    side of a rule in $\asid$ --- each such occurrence of a predicate
    atom $\apred(z_1, \ldots, z_{\arityof{\apred}})$ can be replaced
    by $\apred_0(z_1, \ldots, z_{\arityof{\apred}})$, where $\apred_0$
    is a fresh predicate with the same definition as $\apred$,
  \item\label{ass2} each rule that defines $\apred$ has at least one occurrence of
    a predicate atom --- rules of the form $\apred(x_1, \ldots,
    x_{\arityof{\apred}}) \leftarrow \phi$, where $\phi$ has no
    predicate atoms can be removed without affecting the
    \twoexptime-hardness result from
    \cite[Thm. 4]{BozgaBueriIosif22Arxiv} because only a finite subset
    of the $\asid$-models of $\apred(x_1,\ldots,x_{\arityof{\apred}})$
    (modulo renaming of components) gets lost.
  \end{compactenum}
  We define the following copies of $\asid$: \begin{compactitem}
  \item $\asid_0$ is the following set of rules:
    \begin{align*}
      \apred(x_1, \ldots, x_{\arityof{\apred}}) \leftarrow & \exists y_1 \ldots \exists y_m ~.~ \compactin{x_1}{q_0} * \psi \text{, where} \\
      & \apred(x_1, \ldots, x_{\arityof{\apred}}) \leftarrow \exists y_1 \ldots \exists y_m ~.~ \compact{x_1} * \psi \in \asid \\
      & \text{ and $\psi$ contains no component atoms} \\\\[-2mm]
      \bpred'(x_1, \ldots, x_{\arityof{\bpred'}}) \leftarrow & \exists y_1 \ldots \exists y_m ~.~ \compactin{x_1}{q_1} * \psi \text{, where} \\
      & \bpred'(x_1, \ldots, x_{\arityof{\bpred'}}) \leftarrow \exists y_1 \ldots \exists y_m ~.~ \compact{x_1} * \psi \in \asid, \\
      & \apred \neq \bpred' \text{ and $\psi$ contains no component atoms}
    \end{align*}
  \item $\asid_1$ is the following set of rules:
    \begin{align*}
      \bpred'(x_1, \ldots, x_{\arityof{\bpred}}) \leftarrow & \exists y_1 \ldots \exists y_m ~.~ \compactin{x_1}{q_1} * \psi \text{, where} \\
      & \bpred'(x_1, \ldots, x_{\arityof{\bpred}}) \leftarrow \exists y_1 \ldots \exists y_m ~.~ \compact{x_1} * \psi \in \asid, \\
      & \text{ and $\psi$ contains no component atoms}
    \end{align*}
  \end{compactitem}
  Let $\overline{\asid}$ be the union of $\asid_0$ and $\asid_1$ to
  which the following rules are added, for a fresh predicate
  $\overline{\apred}$ of the same arity as $\apred$:
  \begin{align*}
    \overline{\apred}(x_1, \ldots, x_{\arityof{\apred}}) \leftarrow & \exists y_1 \ldots \exists y_m ~.~ \compactin{x_1}{q} * \psi \text{, where} \\
    & \apred(x_1, \ldots, x_{\arityof{\apred}}) \leftarrow \exists y_1 \ldots \exists y_m ~.~ \compactin{x_1}{q} * \psi \in \asid_0 \\\\[-2mm]
    \overline{\apred}(x_1, \ldots, x_{\arityof{\apred}}) \leftarrow & \exists x_{\arityof{\apred}+1} \ldots \exists x_{\arityof{\bpred}}
    \exists y_1 \ldots \exists y_m ~.~ \compactin{x_1}{q} * \psi \text{, where} \\
    &  \bpred(x_1, \ldots, x_{\arityof{\bpred}}) \leftarrow \exists y_1 \ldots \exists y_m ~.~ \compactin{x_1}{q} * \psi \in \asid
  \end{align*}
  We prove that $\hinv{\overline{\asid}}{\overline{\apred}}$ has a
  positive answer if and only if $\apred(x_1, \ldots,
  x_{\arityof{\apred}}) \models_\asid \exists x_{\arityof{\apred}+1}
  \ldots \exists x_{\arityof{\bpred}} ~.~ \bpred(x_1, \ldots,
  x_{\arityof{\bpred}})$.

  \vspace*{\baselineskip}
  \noindent``$\Rightarrow$'' Let $\aconfig = \config$ be a
  configuration and $\store$ be a store such that $\aconfig
  \models_\asid^\store \apred(x_1, \ldots, x_{\arityof{\apred}})$.
  Let $\aconfig_0 = (\comps, \interacs, \statemap_0)$ be the
  configuration such that $\statemap_0(\store(x_1)) = q_0$ and
  $\statemap_0(c)=q_1$, for all $c \in \comps \setminus
  \set{\store(x_1)}$. By the definition of $\asid_0$, we have
  $\aconfig_0 \models_{\asid_0}^\store \apred(x_1, \ldots,
  x_{\arityof{\apred}})$, hence also $\aconfig_0
  \models_{\overline{\asid}}^\store \overline{\apred}(x_1, \ldots,
  x_{\arityof{\apred}})$. Let $\aconfig_1 \isdef (\comps, \interacs,
  \statemap_1)$, where $\statemap_1 = \statemap_0[\store(x_1)
    \leftarrow q_1]$. Since $\overline{\asid}$ is progressing and
  connected, by the assumption (\ref{ass2}) above, there exists an
  interaction involving $\store(x_1)$, then $\aconfig_1$ is the result
  of executing that interaction in $\aconfig_0$, hence $\aconfig_0
  \Havoc \aconfig_1$. By the hypothesis
  $\hinv{\overline{\asid}}{\overline{\apred}}$, we obtain $\aconfig_1
  \models \overline{\apred}(x_1, \ldots, x_{\arityof{\apred}})$ and,
  since $\statemap_1(c) = q_1$ for all $c \in \comps$, it must be the
  case that $\aconfig_1 \models_{\asid_1}^{\store} \exists
  x_{\arityof{\apred}+1} \ldots \exists x_{\arityof{\bpred}} ~.~
  \bpred(x_1, \ldots, x_{\arityof{\bpred}})$. Since no state atoms
  occur in $\asid$, we obtain $\aconfig \models_\asid \exists
  x_{\arityof{\apred}+1} \ldots \exists x_{\arityof{\bpred}} ~.~
  \bpred(x_1, \ldots, x_{\arityof{\bpred}})$, by the definition of
  $\asid_1$.

  \vspace*{\baselineskip}
  \noindent``$\Leftarrow$'' Let $\aconfig = \config$ and $\aconfig' =
  (\comps, \interacs, \statemap')$ be configurations and $\store$ be a
  store such that $\aconfig \models_{\overline{\asid}}^\store
  \overline{\apred}(x_1, \ldots, x_{\arityof{\apred}})$ and $\aconfig
  \Havoc \aconfig'$. If $\aconfig=\aconfig'$ there is nothing to
  prove. Otherwise, the only possibility is that
  $\statemap(\store(x_1))=q_0$ and $\statemap'(\store(x_1))=q_1$. Then
  we have $\aconfig \models_{\asid_0}^\store \apred(x_1, \ldots,
  x_{\arityof{\apred}})$, by the definition of $\overline{\asid}$ and
  $\aconfig \models_{\asid}^\store \apred(x_1, \ldots,
  x_{\arityof{\apred}})$ follows from the assumption that there are no
  state atoms in $\asid$. By the hypothesis, we obtain $\aconfig
  \models_{\asid}^\store \exists x_{\arityof{\apred}+1} \ldots \exists
  x_{\arityof{\bpred}} ~.~ \bpred(x_1, \ldots, x_{\arityof{\bpred}})$,
  hence $\aconfig' \models_{\asid_1}^\store \exists
  x_{\arityof{\apred}+1} \ldots \exists x_{\arityof{\bpred}} ~.~
  \bpred(x_1, \ldots, x_{\arityof{\bpred}})$, by the definition of
  $\asid_1$, leading to $\aconfig' \models_{\overline{\asid}}
  \overline{\apred}(x_1, \ldots, x_{\arityof{\apred}})$. 

  We conclude observing that the construction of $\overline{\asid}$
  takes time linear in the size of $\asid$.
\end{proofE}

The main result of this section is a consequence of Theorems
\ref{thm:reduction} and \ref{thm:entailment}. In the absence of a
constant bound on the parameters $\maxarityof{\asid}$,
$\maxinterof{\asid}$ and $\maxpredsof{\asid}$, the entailment
resulting from the reduction (Theorem \ref{thm:reduction}) is of
simply exponential size in the input and the time complexity of
solving the entailments is \fourexptime (Theorem
\ref{thm:entailment}), yielding a \fiveexptime\ upper bound:

\begin{theoremE}\label{thm:complexity}
  The $\hinv{\asid}{\apred}$ problem, for \textsf{PCR} SIDs such that
  $\bnd{\asid}{\apred}$ has a positive answer is in \twoexptime, if
  $\maxarityof{\asid}$, $\maxinterof{\asid}$ and $\maxpredsof{\asid}$
  are bounded by constants and in \fiveexptime, otherwise.
\end{theoremE}
\begin{proofE}
  If $\maxarityof{\asid}$, $\maxinterof{\asid}$ and
  $\maxpredsof{\asid}$ are bounded by constants, each instance
  $\hinv{\asid}{\apred}$ of the havoc invariance problem can be
  converted, in polynomial time, into several instances
  $\Set{\overline{\apred}_i(x_1, \ldots, x_{\arityof{\apred}})
    \models_{\asid \cup \overline{\asid}} \apred(x_1, \ldots,
    x_{\arityof{\apred}})}_{i=1}^p$ of the entailment problem, such
  that $\overline{\apred}_i \sim \apred$, for all $i \in
  \interv{1}{p}$ and $\asid \simeq \overline{\asid}$, by Theorem
  \ref{thm:reduction}. Hence $\overline{\asid}$ is \textsf{PCR} and
  $\bnd{\overline{\asid}}{\overline{\apred}_i}$ has a positive answer,
  for each $i \in \interv{1}{p}$. By Theorem \ref{thm:entailment}, the
  entailment problems can be answered in \twoexptime\ for each of
  these instances, hence $\hinv{\asid}{\apred}$ can be answered in
  \twoexptime, as well. Otherwise, if either $\maxarityof{\asid}$,
  $\maxinterof{\asid}$ or $\maxpredsof{\asid}$ are unbounded, the
  entailments are produced in time simply exponential (Theorem
  \ref{thm:reduction}) and can be answered in \fourexptime\ in their
  size (Theorem \ref{thm:entailment}), yielding a \fiveexptime\ upper
  bound.
\end{proofE}

\section{Conclusions}

We have considered a logic for describing sets of configurations of
parameterized concurrent systems, with user-defined network
topology. The havoc invariance problem asks whether a given formula in
the logic is invariant under the execution of the system starting from
each configuration that is a model of a formula. An algorithm for this
problem uses a many-one reduction to the entailment problem, thus
leveraging from earlier results on the latter problem. We study the
decidability and complexity of the havoc invariance problem and show
that a doubly-exponential algorithm exists for a fairly general
fragment of the logic, that encompasses all our examples. This result
is relevant for automating the generation of correctness proofs for
reconfigurable systems, that change the network topology at runtime.


\bibliographystyle{abbrv}
\bibliography{refs}

\begin{thebibliography}{10}

\bibitem{AhrensBozgaIosifKatoen21}
Emma Ahrens, Marius Bozga, Radu Iosif, and Joost{-}Pieter Katoen.
\newblock Local reasoning about parameterized reconfigurable distributed
  systems.
\newblock {\em CoRR}, abs/2107.05253, 2021.

\bibitem{BarHillel61}
Yehoshua Bar-Hillel, Micha Perles, and Eli Shamir.
\newblock On formal properties of simple phrase structure grammars.
\newblock {\em Sprachtypologie und Universalienforschung}, 14:143--172, 1961.

\bibitem{BloemJacobsKhalimovKonnovRubinVeithWidder15}
Roderick Bloem, Swen Jacobs, Ayrat Khalimov, Igor Konnov, Sasha Rubin, Helmut
  Veith, and Josef Widder.
\newblock {\em Decidability of Parameterized Verification}.
\newblock Synthesis Lectures on Distributed Computing Theory. Morgan {\&}
  Claypool Publishers, 2015.

\bibitem{BozgaBueriIosif22Arxiv}
Marius Bozga, Lucas Bueri, and Radu Iosif.
\newblock Decision problems in a logic for reasoning about reconfigurable
  distributed systems.
\newblock {\em arXiv 2202.09637, cs.LO}, 2022.

\bibitem{DBLP:conf/tacas/BozgaEISW20}
Marius Bozga, Javier Esparza, Radu Iosif, Joseph Sifakis, and Christoph Welzel.
\newblock Structural invariants for the verification of systems with
  parameterized architectures.
\newblock In {\em Tools and Algorithms for the Construction and Analysis of
  Systems - 26th International Conference, {TACAS} 2020}, volume 12078 of {\em
  LNCS}, pages 228--246. Springer, 2020.

\bibitem{DBLP:conf/facs2/BozgaI21}
Marius Bozga and Radu Iosif.
\newblock Specification and safety verification of parametric hierarchical
  distributed systems.
\newblock In {\em Formal Aspects of Component Software - 17th International
  Conference, {FACS} 2021, Virtual Event, October 28-29, 2021, Proceedings},
  volume 13077 of {\em Lecture Notes in Computer Science}, pages 95--114.
  Springer, 2021.

\bibitem{BucchiaroneG08}
Antonio Bucchiarone and Juan~P. Galeotti.
\newblock Dynamic software architectures verification using dynalloy.
\newblock {\em Electron. Commun. Eur. Assoc. Softw. Sci. Technol.}, 10, 2008.

\bibitem{CalcagnoOHearnYan07}
Cristiano Calcagno, Peter~W. O'Hearn, and Hongseok Yang.
\newblock Local action and abstract separation logic.
\newblock In {\em 22nd {IEEE} Symposium on Logic in Computer Science {(LICS}
  2007), 10-12 July 2007, Wroclaw, Poland, Proceedings}, pages 366--378. {IEEE}
  Computer Society, 2007.
\newblock \href {https://doi.org/10.1109/LICS.2007.30}
  {\path{doi:10.1109/LICS.2007.30}}.

\bibitem{DBLP:conf/concur/CookHOPW11}
Byron Cook, Christoph Haase, Jo{\"{e}}l Ouaknine, Matthew~J. Parkinson, and
  James Worrell.
\newblock Tractable reasoning in a fragment of separation logic.
\newblock In {\em {CONCUR}}, volume 6901 of {\em Lecture Notes in Computer
  Science}, pages 235--249. Springer, 2011.

\bibitem{DormoyKL10}
Julien Dormoy, Olga Kouchnarenko, and Arnaud Lanoix.
\newblock Using temporal logic for dynamic reconfigurations of components.
\newblock In Lu{\'{\i}}s~Soares Barbosa and Markus Lumpe, editors, {\em Formal
  Aspects of Component Software - 7th International Workshop, {FACS} 2010},
  volume 6921 of {\em Lecture Notes in Computer Science}, pages 200--217.
  Springer, 2010.

\bibitem{DBLP:conf/sac/El-HokayemBS21}
Antoine El{-}Hokayem, Marius Bozga, and Joseph Sifakis.
\newblock A temporal configuration logic for dynamic reconfigurable systems.
\newblock In Chih{-}Cheng Hung, Jiman Hong, Alessio Bechini, and Eunjee Song,
  editors, {\em {SAC} '21: The 36th {ACM/SIGAPP} Symposium on Applied
  Computing, Virtual Event, Republic of Korea, March 22-26, 2021}, pages
  1419--1428. {ACM}, 2021.
\newblock \href {https://doi.org/10.1145/3412841.3442017}
  {\path{doi:10.1145/3412841.3442017}}.

\bibitem{DBLP:journals/sigact/FoersterS19}
Klaus{-}Tycho Foerster and Stefan Schmid.
\newblock Survey of reconfigurable data center networks: Enablers, algorithms,
  complexity.
\newblock {\em {SIGACT} News}, 50(2):62--79, 2019.

\bibitem{Hirsch}
Dan Hirsch, Paolo Inverardi, and Ugo Montanari.
\newblock Graph grammars and constraint solving for software architecture
  styles.
\newblock In {\em Proceedings of the Third International Workshop on Software
  Architecture}, ISAW '98, page 69–72, New York, NY, USA, 1998. Association
  for Computing Machinery.
\newblock \href {https://doi.org/10.1145/288408.288426}
  {\path{doi:10.1145/288408.288426}}.

\bibitem{KestenPnueliShaharZuck02}
Yonit Kesten, Amir Pnueli, Elad Shahar, and Lenore~D. Zuck.
\newblock Network invariants in action.
\newblock In {\em {CONCUR} 2002 - Concurrency Theory, 13th International
  Conference}, volume 2421 of {\em LNCS}, pages 101--115. Springer, 2002.

\bibitem{LanoixDK11}
Arnaud Lanoix, Julien Dormoy, and Olga Kouchnarenko.
\newblock Combining proof and model-checking to validate reconfigurable
  architectures.
\newblock {\em Electron. Notes Theor. Comput. Sci.}, 279(2):43--57, 2011.

\bibitem{LeMetayer}
Daniel Le~Metayer.
\newblock Describing software architecture styles using graph grammars.
\newblock {\em IEEE Transactions on Software Engineering}, 24(7):521--533,
  1998.
\newblock \href {https://doi.org/10.1109/32.708567}
  {\path{doi:10.1109/32.708567}}.

\bibitem{LesensHalbwachsRaymond97}
David Lesens, Nicolas Halbwachs, and Pascal Raymond.
\newblock Automatic verification of parameterized linear networks of processes.
\newblock In {\em The 24th {ACM} {SIGPLAN-SIGACT} Symposium on Principles of
  Programming Languages}, pages 346--357. {ACM} Press, 1997.

\bibitem{DBLP:journals/comsur/Noormohammadpour18}
Mohammad Noormohammadpour and Cauligi~S. Raghavendra.
\newblock Datacenter traffic control: Understanding techniques and tradeoffs.
\newblock {\em {IEEE} Commun. Surv. Tutorials}, 20(2):1492--1525, 2018.

\bibitem{Reynolds02}
John~C. Reynolds.
\newblock Separation logic: {A} logic for shared mutable data structures.
\newblock In {\em 17th {IEEE} Symposium on Logic in Computer Science {(LICS}
  2002), 22-25 July 2002, Copenhagen, Denmark, Proceedings}, pages 55--74.
  {IEEE} Computer Society, 2002.
\newblock \href {https://doi.org/10.1109/LICS.2002.1029817}
  {\path{doi:10.1109/LICS.2002.1029817}}.

\bibitem{ShtadlerGrumberg89}
Ze'ev Shtadler and Orna Grumberg.
\newblock Network grammars, communication behaviors and automatic verification.
\newblock In Joseph Sifakis, editor, {\em Automatic Verification Methods for
  Finite State Systems, International Workshop}, volume 407 of {\em LNCS},
  pages 151--165. Springer, 1989.

\bibitem{WolperLovinfosse89}
Pierre Wolper and Vinciane Lovinfosse.
\newblock Verifying properties of large sets of processes with network
  invariants.
\newblock In {\em Automatic Verification Methods for Finite State Systems,
  International Workshop}, volume 407 of {\em LNCS}, pages 68--80. Springer,
  1989.

\end{thebibliography}

\end{document}